\documentclass[traditabstract,longauth]{aa}

\usepackage{txfonts}
\usepackage{natbib}
\usepackage{graphicx}
\usepackage[breaklinks]{hyperref}
\bibpunct{(}{)}{;}{a}{}{,} 

\begin{document}

\title{Identification of $\gamma$-ray emission from 3C$\,$345 and NRAO$\,$512}

\authorrunning{Schinzel et al.}

\author{
F.~K.~Schinzel$^{(1)}$\thanks{Member of the International
Max Planck Research School (IMPRS) for Astronomy and Astrophysics at the
Universities of Bonn and Cologne.} \and 
K.~V.~Sokolovsky$^{(1,2)\star}$ \and 
F.~D'Ammando$^{(3,4)}$ \and 
T.~H.~Burnett$^{(5)}$ \and
W.~Max-Moerbeck$^{(6)}$ \and 
C.~C.~Cheung$^{(7)}$ \and 
S.~J.~Fegan$^{(8)}$ \and
J.~M.~Casandjian$^{(9)}$ \and
L.~C.~Reyes$^{(10)}$ \and 
M.~Villata$^{(11)}$ \and 
C.~M.~Raiteri$^{(11)}$ \and 
I.~Agudo$^{(12,13)}$ \and 
O.~J.~A.~Bravo~Calle$^{(14)}$ \and 
D.~Carosati$^{(15,16)}$ \and 
R.~Casas$^{(17)}$ \and 
J.~L.~G\'omez$^{(13)}$ \and 
M.~A.~Gurwell$^{(18)}$ \and 
H.~Y.~Hsiao$^{(19)}$ \and
S.~G.~Jorstad$^{(12)}$ \and 
G.~Kimeridze$^{(20)}$ \and 
T.~S.~Konstantinova$^{(14)}$ \and 
E.~N.~Kopatskaya$^{(14)}$ \and 
E.~Koptelova$^{(19)}$ \and 
O.~M.~Kurtanidze$^{(20)}$ \and 
S.~O.~Kurtanidze$^{(20)}$ \and 
V.~M.~Larionov$^{(21,22,14)}$ \and 
E.~G.~Larionova$^{(14)}$ \and 
L.~V.~Larionova$^{(14)}$ \and 
A.~P.~Marscher$^{(12)}$ \and 
D.~A.~Morozova$^{(14)}$ \and 
M.~G.~Nikolashvili$^{(20)}$ \and 
M.~Roca-Sogorb$^{(13)}$ \and 
J.~A.~Ros$^{(23)}$ \and 
L.~A.~Sigua$^{(20)}$ \and 
O.~Spiridonova$^{(24)}$ \and 
I.~S.~Troitsky$^{(14)}$ \and 
V.V.~Vlasyuk$^{(24)}$ \and
A.~P.~Lobanov$^{(1)}$ \and
J.~A.~Zensus$^{(1)}$
}

\institute{
\inst{1}~Max-Planck-Institut f\"ur Radioastronomie, Auf dem H\"ugel 69, 53121 Bonn, Germany\\ 
\inst{2}~Astro Space Center of Lebedev Physical Institute, Profsoyuznaya Str. 84/32, 117997 Moscow, Russia\\ 
\inst{3}~INAF-IASF Palermo, 90146 Palermo, Italy\\ 
\inst{4}~INAF-IASF Roma, 00133 Roma, Italy\\ 
\inst{5}~Department of Physics, University of Washington, Seattle, WA 98195-1560, USA\\ 
\inst{6}~Cahill Center for Astronomy and Astrophysics, California Institute of Technology, Pasadena, CA 91125, USA\\ 
\inst{7}~National Research Council Research Associate, National Academy of Sciences, Washington, DC 20001, resident at Naval Research Laboratory, Washington, DC 20375, USA\\ 
\inst{8}~Laboratoire Leprince-Ringuet, \'Ecole polytechnique, Route de Saclay,
91128 Palaiseau, France\\
\inst{9}~Laboratoire AIM, CEA-IRFU/CNRS/Universit\'e Paris Diderot, Service d'Astrophysique, CEA Saclay, 91191 Gif sur Yvette, France\\ 
\inst{10}~Kavli Institute for Cosmological Physics, University of Chicago, Chicago, IL 60637, USA\\ 
\inst{11}~INAF, Osservatorio Astronomico di Torino, I-10025 Pino Torinese (TO), Italy\\ 
\inst{12}~Institute for Astrophysical Research, Boston University, Boston, MA 02215, USA\\ 
\inst{13}~Instituto de Astrof\'isica de Andaluc\'ia, CSIC, E-18080 Granada, Spain\\  
\inst{14}~Astronomical Institute, St. Petersburg State University, St. Petersburg, Russia\\ 
\inst{15}~EPT Observatories, Tijarafe, La Palma, Spain\\ 
\inst{16}~INAF, TNG Fundación Galileo Galilei, La Palma, Spain\\
\inst{17}~Institut de Ciencies de l'Espai (IEEC-CSIC), Campus UAB, 08193 Barcelona, Spain\\ 
\inst{18}~Harvard-Smithsonian Center for Astrophysics, Cambridge, MA 02138, USA\\ 
\inst{19}~Graduate Institute of Astronomy, National Central University, Jhongli 32054, Taiwan\\  
\inst{20}~Abastumani Observatory, Mt. Kanobili, 0301 Abastumani, Georgia\\ 
\inst{21}~Isaac Newton Institute of Chile, St. Petersburg Branch, St. Petersburg, Russia\\ 
\inst{22}~Pulkovo Astronomical Observatory, Pulkovskoe Chaussee 65/1,
196140 St. Petersburg, Russia\\ 
\inst{23}~Agrupaci\'o Astron\`omica de Sabadell, 08206 Sabadell, Spain\\ 
\inst{24}~Special Astrophysical Observatory, Nizhnij Arkhyz, Karachai-Cherkessian Republic, 369167, Russia\\[12pt]
\email{schinzel@mpifr-bonn.mpg.de} \\
}

\date{Accepted for publication in A\&A (07/07/2011)}

\abstract{
For more than 15 years, since the days of the Energetic Gamma-Ray Experiment
Telescope (EGRET) on board the \textit{Compton Gamma-Ray Observatory} (CGRO;
1991-2000), it has remained an open question why the prominent blazar 3C$\,$345 was not
reliably detected at $\gamma$-ray energies $\geq$\,20\,MeV. Recently a bright $\gamma$-ray
source (0FGL\,J1641.4+3939/1FGL\,J1642.5+3947), potentially associated with 3C$\,$345, was detected
by the Large Area Telescope (LAT) on \textit{Fermi}. Multiwavelength observations from radio bands to X-rays (mainly
GASP-WEBT and \textit{Swift}) of possible counterparts (3C$\,$345, NRAO$\,$512, B3$\,$1640$+$396) 
were combined with 20 months of \textit{Fermi}-LAT monitoring data (August 2008 - April 2010)
to associate and identify the dominating $\gamma$-ray emitting counterpart of
1FGL\,J1642.5+3947. The source 3C$\,$345 is identified as the main contributor for this
$\gamma$-ray emitting region. However, after November 2009 (15 months), a significant
excess of photons from the nearby quasar NRAO$\,$512 started to contribute and thereafter
was detected with increasing $\gamma$-ray activity, possibly adding flux to 1FGL\,J1642.5+3947. 
For the same time period and during the summer of 2010, an increase of radio, optical and 
X-ray activity of NRAO$\,$512 was observed. No $\gamma$-ray emission from B3$\,$1640$+$396 was detected.
}


\maketitle

\section{Introduction}

The quasar \object{3C$\,$345} is
known as a prominent variable source from radio- to
X-ray bands ($\leq$\,10\,keV). It is particularly
bright at radio wavelengths and has an extended radio
structure that is observable from sub-pc to kpc scales which is archetypical for a relativistic
blazar jet \citep[e.g.][]{1999ApJ...521..509L, 1989AJ.....97.1550K}. Its jet has an apparent opening 
angle of $\sim$\,13\degr\ \citep{2009A&A...507L..33P}. 
\citet{1994ApJ...432..103U} concluded that the observed high-energy emission of 
this source is caused by inverse Compton up-scattering of lower energy photons by 
relativistic electrons of the pc-scale radio jet. However, it remained a puzzle 
why 3C$\,$345 was not detected at energies above 20\,MeV, while 3C$\,$279 with similar
apparent properties is a prominent source at $\gamma$-ray energies.

The high-redshift quasar \object{NRAO$\,$512},
0.5\degr\ west of 3C$\,$345, is known as a compact radio source with a two-sided jet 
morphology on kpc scales and with a flat radio
spectrum \citep{1999yCat..41390545K}. The observed VLBI radio structure revealed
a wide apparent jet opening angle of 
$\sim$\,50\degr\ \citep{2009A&A...507L..33P}, which implies a small viewing angle to the line of sight, 
with its emission directly beamed at us.

The Energetic Gamma-Ray Experiment Telescope \citep[EGRET;][]{1993ApJS...86..629T} on board the \textit{Compton Gamma
Ray Observatory} (CGRO) spacecraft did not observe a significant excess of
$\gamma$-ray photons consistent with 3C$\,$345 \citep[3EG;][]{1999ApJS..123...79H}, covering the first four phases of the CGRO
mission (1991-1995). This lack of detectable flux of $\gamma$-rays above 20\,MeV from 3C$\,$345 was attributed to the high
detection threshold of EGRET. A source consistent with the position of 3C$\,$345, labeled as EGR$\,$J1642+3940, was 
found by \citet{2008A&A...489..849C}. This source was seen almost entirely in a single two-week viewing period 
with an excess of $\sim$\,5.8$\sigma$, from
April 23rd - May 7th, 1996 (vp5190), which was after the data used for the 3EG. 
Unfortunately, the angular resolution of EGRET was too poor to distinguish contributions from the other two nearby candidates
NRAO$\,$512 and B3$\,$1640$+$396, see Table~\ref{tab:sources}. Between October 1st and October 4th, 1993, a similar source was marginally detected by 
EGRET ($\sim$\,2.1$\sigma$), but its centroid was localized closer to Mrk$\,$501. In exactly the same viewing period,
\citet{1999ApJ...514..138K} claimed EGRET detected a flare from Mrk$\,$501, which lies 2.5\degr\ east of 3C\,345.
This would then be the only time that either of 
those sources was seen by EGRET -- a strange coincidence that led Casandjian \& Grenier to note that the EGR$\,$J1642+3940 
source association with 3C$\,$345 is ``not clear''.

On June 11th, 2008, the \textit{Fermi} Gamma-ray Space Telescope (short
\textit{Fermi}) was launched, equipped
with the Large Area Telescope (LAT) instrument
\citep{2009ApJ...697.1071A} providing continuous all-sky monitoring of
$\gamma$-ray emission as a default mode of operation. The LAT, compared with EGRET,
has an improved sensitivity by a factor of $\sim$\,25 and an improved point-source 
localization. After 20 months of LAT operation within the
region of interest, a localization improvement (comparing radii of 95\% error circles) of a factor of $\sim$\,14 was
reached compared to EGRET. These improvements provided a new opportunity to
probe 3C$\,$345 for $\gamma$-ray emission above 100\,MeV.

The three-month bright \textit{Fermi} source list (0FGL) for the dataset collected between
August and October 2008 reported a $\gamma$-ray source (0FGL\,J1641.4+3939) near
the position of 3C$\,$345. It was then associated
with the quasar B3$\,$1640$+$396 (CLASS$\,$J1641+3935) at low confidence \citep{2009ApJ...700..597A}. The bright
0FGL source was listed in the first \textit{Fermi}-LAT source catalog (1FGL) as
1FGL\,J1642.5+3947 \citep{2010ApJS..188..405A}, further constraining its
localization. An association with 3C$\,$345 was still not possible with
high confidence based on \textit{Fermi}-LAT data alone, nevertheless,
1FGL\,J1642.5+3947 was listed in the catalog as being affiliated with its most likely counterpart 3C$\,$345
\citep{2010ApJ...715..429A}.

In October 2009 a GeV flare was detected from the vicinity of 0FGL\,J1641.4+3939
\citep{2009ATel.2226....1R}. Simultaneously, the GASP-WEBT collaboration
reported on increased optical and radio activity of
3C$\,$345 \citep{2009ATel.2222....1L}. This event triggered additional observations 
at different frequencies, which made it possible to establish
an identification through correlated multiwavelength activity. 

Here we present 20 months of monitoring observations by \textit{Fermi} LAT of the
region around 3C$\,$345, NRAO$\,$512 and B3$\,$1640$+$396, as well as radio, optical, 
UV, and X-ray observations of these sources.
Section 2 discusses the observations and data reduction methods applied to
obtain calibrated datasets for the subsequent analysis. Section 3 discusses the results
of \textit{Fermi}-LAT observations and identification through multiwavelength
counterparts of events in the light curve. Section 4 discusses and summarizes the findings of this paper.

\section{Observations and data reduction methods}

\begin{table}
   \caption{Radio positions and redshifts of candidate $\gamma$-ray counterparts.}
   \label{tab:sources}
   \centering
   \begin{tabular}{llll}
   \hline\hline
   Name(s) & \object{3C\,345} & \object{NRAO\,512} & \object{B3$\,$1640$+$396} \\
	   &  &                                    & \object{CLASS$\,$J1641+3935} \\
   \hline
   R.A.    & 16$^\mathrm{h}$42$^\mathrm{m}$58\fs810 & 16$^\mathrm{h}$40$^\mathrm{m}$29\fs633 & 16$^\mathrm{h}$41$^\mathrm{m}$47\fs540\\
   Dec.    & +39\degr48\arcmin36\farcs99            & +39\degr46\arcmin46\farcs03            & +39\degr35\arcmin03\farcs33\\
   z       & 0.5928                                 & 1.666                                  & 0.540\\
   Refs.   & (1, 3) & (1, 4) & (2, 5) \\
   \hline
   \end{tabular}
   \begin{flushleft}
    \small
    Refs. -- The numerals in the references row translate to:\\
	      (1) \citet{2004AJ....127.3587F}; (2) \citet{2003MNRAS.341....1M};
	     (3) \citet{1996ApJS..104...37M}; (4) \citet{1989A&AS...80..103S}; (5) \citet{1995A&AS..109..147B}\\[6pt]
    {\bf Notes:}
    Name(s) -- Common source names.\\
    R.A./Dec. -- Right ascension/declination (J2000) of the radio source localization. \\
    z -- Source redshift.
  \end{flushleft}
\end{table}

\subsection{$\gamma$-ray monitoring}

	The \textit{Fermi} LAT \citep{2009ApJ...697.1071A} is a pair-conversion
telescope designed to cover the energy range from 20\,MeV to greater than
300\,GeV. The 1FGL catalog, based on 11 months of monitoring
data, lists the $\gamma$-ray source \object{1FGL\,J1642.5$+$3947} \citep[R.A. 16$^\mathrm{h}$42$^\mathrm{m}$33.55$^\mathrm{s}$, 
Dec. $+$39\degr47\arcmin38.8\arcsec, 95\% position error radius r95$=$0.056\degr;][]{2010ApJS..188..405A}, which lies in close
vicinity of the three quasars 3C$\,$345, NRAO$\,$512 and B3$\,$1640$+$396, listed in Table~\ref{tab:sources}. 

	The $\textit{Fermi}$-LAT data on the field in which 1FGL\,J1642.5$+$3947 is
located were processed and analyzed using the \textit{Fermi} Science Tools software
package \citep[version: v9r15p2\footnote{\href{http://fermi.gsfc.nasa.gov/ssc/data/analysis/documentation/Cicerone/}{http://fermi.gsfc.nasa.gov/.../documentation/Cicerone}};][]{2010ApJS..188..405A}. 
The photons were extracted from a region of interest (ROI) centered on the
radio position of 3C$\,$345 within a radius of 15$^\circ$ and a time range of 20 months, 
August 4th, 2008, to April 4th, 2010 (in MET\footnote{Mission Elapsed Time (MET) is measured in seconds from
2001.0 UT, including leap seconds.} 239557417 - 292088627\,s). The general data
processing procedures applied were similar to those described in
\citet{2009ApJ...700..597A} and \citet{2010ApJS..188..405A}.

To ensure that the collected events considered in this analysis have a high probability of 
being photons, the ``diffuse class'' selection was applied. Furthermore, events above the zenith angle
105\degr\ were removed to avoid a significant contamination by $\gamma$ rays produced from cosmic-ray
interactions in Earth's atmosphere\footnote{The Earth's limb
lies at a zenith angle of $\sim$113\degr\ at the 565 km
nearly circular orbit of \textit{Fermi}. The zenith angle is defined as the angle of a photon's apparent origin to 
the Earth-spacecraft vector.} \citep{PhysRevD.80.122004}. The 
P6\_V3\_DIFFUSE instrument response functions (IRF) were used \citep[see][]{2009arXiv0907.0626R}.

The remaining $\gamma$-ray-like events were analyzed using a maximum-likelihood 
approach \citep{1996ApJ...461..396M} to localize $\gamma$-ray sources
and extract their spectra. As in the analysis of 1FGL, two sets of
standard LAT analysis tools were used, and their results compared for
consistency.  The first set, based on \texttt{gtfindsrc} and
\texttt{gtlike} tools, employs an unbinned likelihood analysis in
which the expected response of the LAT is evaluated separately for
each photon, given its energy and direction with respect to the
detector axes. The second toolset, \texttt{pointlike}, uses a simpler,
binned analysis methodology, where photons are binned spatially
and in energy, and an averaged set of response functions is applied.
The two packages give very similar results, with the
\texttt{pointlike} package running much more quickly.

In both likelihood methodologies the $\gamma$-ray emission in the ROI is
assumed to arise from two diffuse components and a number of point
sources. Each point source is modeled spatially as a delta function
and spectrally as a power law. The statistical significance of a
potential source is estimated by forming a ``test statistic'' (TS),
\begin{equation}
\mathrm{TS} = 2\,(\ln L_1 - \ln L_0),
\end{equation}
where $L_1$ is the value of the likelihood obtained in the ``source''
hypothesis, with the source present in the model of the ROI and $L_0$
is the value in the alternate, ``null'' hypothesis, that the source is
not real (and hence not included in the ROI model). If the null
hypothesis is true, and assuming certain applicability conditions are
met, Wilks' theorem \citep{1938AMS...9...60W} states that TS is distributed as
$\chi^{2}\left(N\right)$, where $N$ is the number of additional parameters
optimized in the source model leading to $L_1$. Because each point
source has four free parameters (two for its position and two for its
spectrum), we expect that $TS\sim\chi^{2}\left(4\right)$, from which its
significance can be evaluated. A source-detection threshold of TS=25
was adopted in 1FGL, which corresponds to a false detection
probability of $5\times10^{-5}$ if $TS\sim\chi^{2}\left(4\right)$.

The \texttt{gtlike} model used for localization in this paper includes the 1FGL\,J1642.5$+$3947 point-source component 
and all other point sources from the 1FGL catalog \citep{2010ApJ...715..429A} within a
20\degr\ radius around the radio position of 3C$\,$345. Further analysis was performed using two point-source components
instead of one to fit the observed data. The model parameters for sources outside the
15\degr\ ROI were fixed to their respective 1FGL values. The model included 20 or 21 point sources with free parameters 
(flux and spectral index), depending on whether a single point source was fitted for 1FGL\,J1642.5$+$3947 (20 point sources) or whether
 two source components were added to take into account the separate contributions of 3C$\,$345 and NRAO$\,$512 (21 point sources). 
The background component of Galactic diffuse emission (\texttt{gll\_iem\_v02.fit}) 
and an isotropic component (\texttt{isotropic\_iem\_v02.txt}), both of which are the standard models available from the 
\textit{Fermi} Science Support Center (FSSC)\footnote{
\href{http://fermi.gsfc.nasa.gov/ssc/data/access/lat/BackgroundModels.html}{http://fermi.gsfc.nasa.gov/.../lat/BackgroundModels.html}}, were added as well. The
isotropic component includes both the contribution from the extragalactic
diffuse emission and from the residual, charged particle backgrounds.

To produce light curves, the standard tool \texttt{gtlike} was used, applying an unbinned spectral likelihood analysis as
described above. A \texttt{gtlike} input source model was constructed based on
the improved 20-month spectral fits obtained from the localization analysis. Only sources that were
detected with a TS\,$>$\,25 in the 20-month integration were
included in the final \texttt{gtlike} input source model to construct a light
curve. The data were split into regular time intervals, each integrating over periods of 2, 5 and 30\,days. Then an 
unbinned spectral likelihood analysis was performed on each integrated dataset with photon 
indices fixed to the 20-month average values. For time periods with a significance of the detection 
of TS\,$<$\,5 on 5- and 30-day scales, 2$\sigma$ upper limits were calculated following the
likelihood profile \citep{2005NIMA.551.493...R}, where the upper limit is determined
by increasing the flux from the maximum likelihood value until the log likelihood decreases by 2.0 (for the particular
implementation in case of LAT data see \citealt{2010ApJS..188..405A}). All light curves make use of the collected
data in the energy range of 0.1-300\,GeV. Different time binnings were chosen to study different aspects of the 
time evolution of the source flux. To study the longer term behavior, 30-day bins were used (Section~\ref{sec:long}), 
monthly bins being relatively standard for \textit{Fermi} analysis. To assess faster flaring behavior, and in particular to estimate 
when the peak flaring occurs, a two-day time binning was selected (Section~\ref{sec:short}). Finally, the best compromise between short 
and long time bins is the light curve with five-day integration periods (Section~\ref{sec:3.3}).

\subsection{X-ray observations}

Triggered by the reported $\gamma$-ray flare of 0FGL\,J1641.4+3939 in October
2009 \citep{2009ATel.2226....1R}, X-ray observations of all three candidate
counterpart sources were performed by the X-ray Telescope
\citep[XRT, 0.2$-$10\,keV;][]{2005SSRv..120..165B} on board the \textit{Swift}
satellite \citep{2004ApJ...611.1005G}. The list of \textit{Swift}-XRT 
observations together with the spectral fit results is reported in Table~\ref{tab:XRTdata}.

The XRT data were processed with standard procedures (\texttt{xrtpipeline}
v0.12.4) to calibrate the observations. The filtering and screening 
criteria were applied by means of the FTOOLS in the \textit{Heasoft} package version 6.8\footnote{\texttt{xrtpipeline}
and FTOOLS are part of the \textit{Heasoft} software package: \href{http://heasarc.gsfc.nasa.gov/lheasoft/}{http://heasarc.gsfc.nasa.gov/lheasoft/}; a detailed
discussion on XRT data analysis can be found in the XRT User's Guide: \href{http://heasarc.gsfc.nasa.gov/docs/swift/analysis/xrt_swguide_v1_2.pdf}{http://heasarc.gsfc.nasa.gov/.../xrt\_swguide\_v1\_2.pdf}}.
Given the low fluxes of the sources during observations ($<$\,0.5\,counts\,s$^{-1}$ 
in the 0.2$-$10\,keV range), only photon counting
(PC) data were considered for our analysis with XRT grades 0$-$12 selected
\citep[according to \textit{Swift} nomenclature, see][]{2005SSRv..120..165B}. 
The source events were extracted in a circle of
radius 15$-$20 pixels (covering 85-90\% of the XRT point spread function, PSF, at 1.5\,keV)
around the source, depending on its intensity. The background was estimated from
a nearby source-free circular region of 40-pixel radius. Ancillary response files were generated with {\tt
xrtmkarf} to account for different extraction regions, vignetting and PSF
corrections. The spectral redistribution matrices v011 in the
calibration database maintained by HEASARC were used.

The adopted energy range for spectral fitting is 0.3$-$10\,keV. All data were
rebinned with a minimum of 20 counts per energy bin to use the
$\chi^{2}$ minimization fitting technique, except when the statistics were poor ($<$ 200 counts),
in which case the Cash statistic \citep{1979ApJ...228..939C} was used and
the data were binned to 1 counts bin$^{-1}$. The Cash
statistic is based on a likelihood ratio test and is widely used for parameter estimation in 
photon counting experiments. \textit{Swift}-XRT uncertainties are given at
90$\%$ confidence level for each parameter, unless stated otherwise.

Spectral fitting was performed using \textit{XSPEC} version 12.5.1 with an
absorbed power law model. For the Galactic absorption a fixed \ion{H}{I} column
density n$_\mathrm{H}$ of $1.14\cdot10^{20}$~cm$^{-2}$ was used for 3C$\,$345,
$1.17\cdot10^{20}$~cm$^{-2}$ for B3$\,$1640$+$396 and
$1.07\cdot10^{20}$~cm$^{-2}$ for NRAO$\,$512, obtained from the
Leiden/Argentine/Bonn Galactic \ion{H}{I} survey \citep{2005A&A...440..775K}.

 \begin{table*}
   \caption{\textit{Swift}-XRT observations after the reported flare in
 October 2009 of the three candidate sources (3C$\,$345, NRAO$\,$512 and
B3$\,$1640$+$396). \label{tab:XRTdata}}
   \centering
   \begin{tabular}{llccc}
   \hline\hline
   Source & Obs. Date (*) & Exp. & Flux & Index \\
   \hline
   3C$\,$345: & 2009-10-04 (5109)  &  4.2 & 6.8 $\pm$ 0.5 & 1.68 $\pm$ 0.11\\
           & 2009-10-06$^{a}$ (5111)  &  1.9 & 4.7 $\pm$ 0.7 & 1.66 $\pm$ 0.17\\
        & 2009-10-08 (5113)  &  2.1 & 7.1 $\pm$ 0.8 & 1.75 $\pm$ 0.16\\
        & 2009-10-09 (5114)  &  2.2 & 6.4 $\pm$ 0.7 & 1.62 $\pm$ 0.15\\
        & 2009-10-18 (5123)  &  3.9 & 6.2 $\pm$ 0.5 & 1.67 $\pm$ 0.12\\
        & 2009-11-01 (5137)  &  4.1 & 5.5 $\pm$ 0.5 & 1.77 $\pm$ 0.11\\
        & 2009-11-15 (5151)  &  4.0 & 5.9 $\pm$ 0.5 & 1.88 $\pm$ 0.11\\
        & 2009-11-29 (5165)  &  4.1 & 5.7 $\pm$ 0.5 & 1.71 $\pm$ 0.12\\
        & 2010-03-06/07 (5262/3) &2.6 & 5.6 $\pm$ 0.6 & 1.84 $\pm$ 0.15\\
        & 2010-03-09$^{a}$ (5265)  &  1.4 & 5.6 $\pm$ 0.8 & 1.78 $\pm$ 0.18\\
        & 2010-04-09$^{a}$ (5296)  &  2.0 & 5.5 $\pm$ 0.7 & 1.88 $\pm$ 0.16\\
        & 2010-08-18 (5427)  &  4.3 & 5.4 $\pm$ 0.5 & 1.54 $\pm$ 0.11\\
  NRAO$\,$512:& 2009-10-08$^{a}$ (5113)  & 2.1 & 0.6 $^{+0.3}_{-0.2}$ & 1.5
 $\pm$ 0.5\\
        & 2009-10-15$^{a}$ (5120)   &  2.1 & 0.5 $\pm$ 0.2 & 2.2 $\pm$ 0.6\\
	& 2010-08-06/07$^{a}$ (5415/6)   &  5.5 & 0.7 $\pm$ 0.1 & 2.0 $\pm$
0.3\\
   B3$\,$1640+396: & 2009-12-01$^{a}$ (5167) &  1.1 & 0.7 $^{+0.4}_{-0.3}$ &
 1.8 $\pm$ 0.7\\
       \hline
   \end{tabular}
   \begin{flushleft}
     \small
     {\bf Notes:}
     Obs. Date (*) -- Date of \textit{Swift}-XRT observation (*=JD-2450000).\\
     Exp. -- \textit{Swift}-XRT effective exposure time in ks. \\
     Flux -- Integrated unabsorbed photon flux 0.3-10~keV in units
 of 10$^{-12}$~erg~cm$^{-2}$~s$^{-1}$.\\
     $^{a}$ Cash statistic used for observations with less than 200 photons.\\
   \end{flushleft}
 \end{table*}

\subsection{UV, optical and mm observations\label{sec:optical}}

In the case of 3C$\,$345 a dense optical monitoring was performed by the
GLAST-AGILE Support Program (GASP) of the Whole Earth Blazar Telescope (WEBT)
collaboration.
The GASP-WEBT is performing long-term monitoring of
28 $\gamma$-ray loud blazars in the optical, near-infrared, mm and
radio bands \citep{2008A&A...481L..79V, 2009A&A...504L...9V}. The GASP has been
closely following
a strong optical outburst in 3C$\,$345 reported during the second half of
2009 \citep{2009ATel.2222....1L}. The optical GASP data for this paper were
acquired at the following observatories: Abastumani (Georgia), Calar Alto
(Spain, within the MAPCAT program), Crimean (Ukraine), Lowell
(Perkins, USA), Lu-lin (SLT, Taiwan), Sabadell (Spain), St. Petersburg (Russia), and
Tijarafe (Spain).

Magnitude calibration was performed with respect to a common choice of reference
stars in the field of the source obtained from the photometric sequence by
\citet{2001AJ....122.2055G}. In addition to the GASP monitoring data, optical
observations performed by the Special Astrophysical
Observatory (SAO) of the Russian Academy of Sciences, using the SAO Zeiss 1-m
telescope in BVR bands were included. The observational strategy and calibration procedures applied were described in detail 
by \citet{2004ARep...48..840B}. In Section~\ref{sec:3.3} the R band magnitudes
were converted to the linear flux density scale to compare 
light curves between wavelengths. The magnitudes were corrected for 
a Galactic extinction of 0.036 mag, for the magnitude to flux conversion the zero magnitude
fluxes of \citet{1998A&A...333..231B} were used.

Two R-band observations of 3C$\,$345 and B3$\,$1640$+$396 were conducted on
January 26th, 2009
(14x180~s exposure), and on February 13th, 2009 (13x180~s
exposure), using a robotic 200-mm telescope of the Tzec Maun
observatory (Mayhill, NM, USA).
Magnitude calibration was performed in a manner similar to the GASP
observations.

Observations of 3C$\,$345 and NRAO$\,$512 were performed with the
Ultraviolet/Optical Telescope (UVOT) on board the \textit{Swift} spacecraft
\citep{2005SSRv..120...95R}. Optical photometry of 3C$\,$345 was calibrated
using comparison stars from \citet{1985AJ.....90.1184S}. Calibration of the
UV data was performed following the method of \citet{2008MNRAS.383..627P}.
Comparison stars in the vicinity of 3C$\,$345 could not be used to
calibrate optical photometry of NRAO$\,$512 with \textit{Swift} UVOT because
of its narrow field of view. Instead, the magnitude scale was calibrated
using Sloan Digital Sky Survey \citep[SDSS;][]{2009ApJS..182..543A} stars in the
field and transformations between the SDSS
and Johnson photometric systems from \citet{2002AJ....123.2121S}.

Table~\ref{tab:opticaldata} lists the relevant results of
mainly quasi-simultaneous optical/UV observations of 3C$\,$345, NRAO$\,$512 and B3$\,$1640$+$396 performed at Tzec Maun 
observatory and \textit{Swift} UVOT. The Galactic
extinction in the direction of these sources is between E(B$-$V)\,$=$\,0.010 mag and 0.014 mag according to the tables
of \citet{1998ApJ...500..525S}. Using the extinction law by \citet{1989ApJ...345..245C} and coefficients 
describing the UVOT filters \citep{2009ApJ...690..163R}, the following extinctions (in magnitudes) were obtained
for the individual bands: 0.03$-$0.04 (V), 0.04$-$0.06 (B), 0.05$-$0.07 (U), 0.07$-$0.09 (W1), 0.09$-$0.13 (M2) and 0.08$-$0.12 (W2).

The source 3C\,345 was monitored also at mm wavelengths (230 GHz) by the
Sub-millimeter Array (SMA) on Mauna Kea for the GASP. Additionally, a
program conducted by Ann Wehrle (\textit{private communication}) provided the
data collected by the SMA for NRAO 512 and B3 1640+396.

\begin{table}
  \caption{Quasi-simultaneous optical observations of 3C$\,$345, NRAO$\,$512 and
B3$\,$1640$+$396 conducted by \textit{Swift} UVOT and the 200-mm telescope of
Tzec Maun observatory. \label{tab:opticaldata}}
  \centering
  \begin{tabular}{lcccc}
  \hline\hline
  Source & Obs. Date (*) & Band & Magnitude \\
  \hline
  3C$\,$345: & 2009-01-26 (4858) & R & 16.48 $\pm$ 0.02 \\
	  & 2009-02-13 (4876) & R & 16.92 $\pm$ 0.03 \\
	  & 2009-10-08 (5113) & U & 16.44 $\pm$ 0.06 \\
          & 2009-10-08 (5113) & UVW1 & 16.28 $\pm$ 0.03 \\
	  & 2010-08-18 (5427) & UVW1 & 16.47 $\pm$ 0.02 \\
  NRAO$\,$512:& 2009-10-08 (5113) & U & 18.61 $\pm$ 0.04 \\
	   &  2010-08-06 (5415) & U & 17.71 $\pm$ 0.05 \\
	   &  2010-08-06 (5415) & UVW1 & 18.07 $\pm$ 0.02 \\
  B3$\,$1640+396: & 2009-01-26 (4858) & R & 18.57 $\pm$ 0.13 \\
	      & 2009-02-13 (4876) & R & $>18.6^\dagger$\\
  \hline
  \end{tabular}
  \begin{flushleft}
    \small
    {\bf Notes:}
    Obs. Date (*) -- Date of \textit{Swift}-XRT observation (*=JD-2450000).\\
    Band -- Optical filter used for photometric observations\\
    Magnitude -- Observed optical magnitude in given filter band. The magnitudes are not 
    corrected for Galactic extinction.\\
    $\dagger$ no detection\\
  \end{flushleft}
\end{table}

\section{Results}

\subsection{Localization of $\gamma$-ray emission\label{sec:3.1}}

An unbinned spectral likelihood analysis was performed using integrated 20-month
\textit{Fermi}-LAT monitoring data in the energy range of 0.1$-$100\,GeV. This energy range
was chosen after a series of different energy cuts were compared, to minimize the error on 
the localization. In the 1FGL catalog $\gamma$-ray sources are typically localized with a 95\% position error radius 
of 0.02\degr\,--\,0.07\degr\ for sources with TS\,$\approx$\,1000, but 0.06\degr\,--\,0.5\degr\ 
for TS\,$\approx$\,30 \citep{2010ApJS..188..405A}. In cases of faint $\gamma$-ray emitters it is difficult to obtain 
well-constrained localizations to make statistically significant associations, even more so for two 
faint $\gamma$-ray emitters in close vicinity of each other.

A single point-source, 1FGL\,J1642.5+3947, is listed in the 1FGL catalog around the analyzed region. Assuming 
a single point-source, the best-fit position using the 20-month LAT dataset is R.A.\,16$^\mathrm{h}$42$^\mathrm{m}$24$^\mathrm{s}$, Dec.\,+39\degr48\arcmin27\arcsec, 
with a 95\% error circle radius of 0.037\degr. This procedure also provided the following characteristic source 
parameters: photon index\footnote{The photon spectral index $\Gamma$ is defined as N(E)\,$\propto$\,E$^{-\Gamma}$, where N(E) is the
$\gamma$-ray photon flux as a function of energy E.}: $\Gamma\, =\,$2.44$\,\pm\,$0.03, integrated flux: $\left(1.78\,\pm\,0.08\right)\cdot
10^{-7}\,$ph\,cm$^{-2}$\,s$^{-1}$, TS value of 2825. The estimated systematic uncertainty of the integrated 
flux is 10\% at 100\,MeV, 5\% at 500\,MeV and 20\% at 10\,GeV. After 20 months of data the localized position is shifted by 0.042\degr\ from
its 1FGL position toward NRAO\,512, separated by 0.13\degr\ (east) from 3C$\,$345 and by 0.48\degr (west) from NRAO$\,$512. This suggested a significant 
contribution of excess photons from the direction west of 3C$\,$345.   

\begin{figure}
  \includegraphics[width=\hsize]{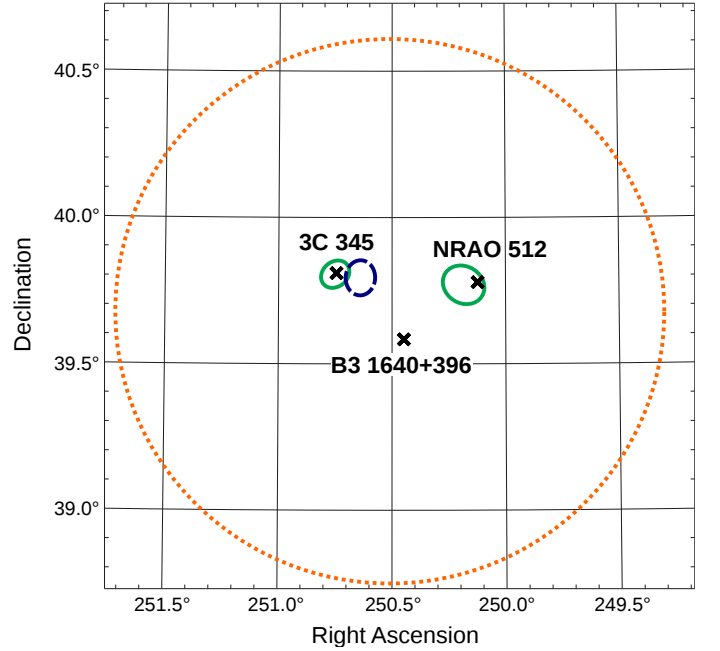}
  \caption{$\gamma$-ray counterpart localizations together with
the positions of the radio counterparts of the three candidate sources. Crosses mark the 
radio positions of 3C$\,$345, NRAO$\,$512 and B3$\,$1640$+$396. The large dotted orange circle denotes the 
95\% confidence error localization of EGR\,J1642+3940 \citep{2008A&A...489..849C}. The dashed blue ellipse 
denotes the 95\% confidence error localization of 1FGL\,J1642.5+3947 \citep{2010ApJS..188..405A}. The two solid green 
ellipses denote the 95\% confidence error localization of 3C$\,$345 and NRAO$\,$512 presented in this 
paper. The confusing source Mrk$\,$501 is 2.5\degr\ east (left) and 4C +38.41 is 2.1\degr\ southwest of the field center.\label{fig:ts_map}}
\end{figure}

\begin{figure}
  \centering
  \includegraphics[width=0.8\hsize]{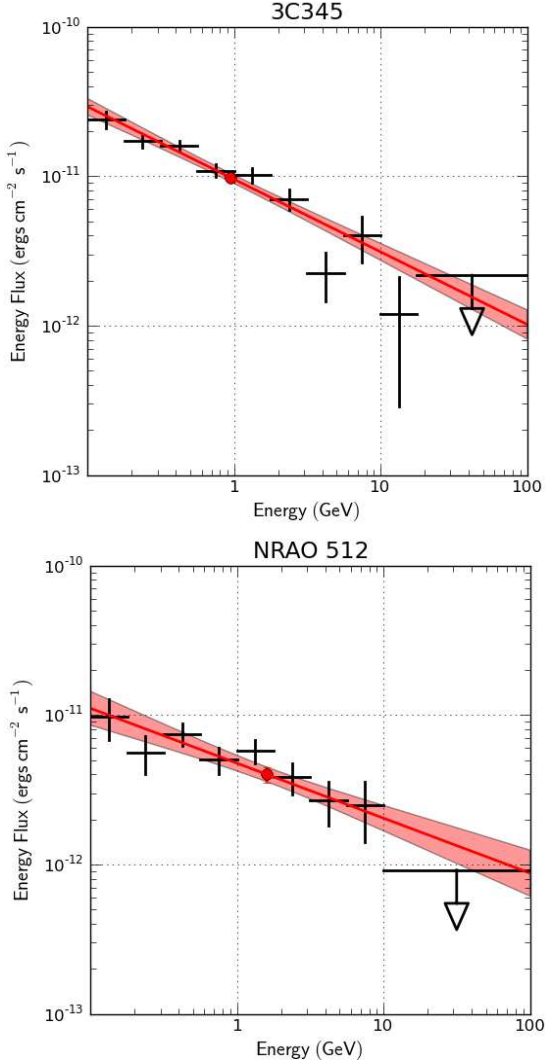}
  \caption{Twenty-month $\gamma$-ray spectra of 3C$\,$345 and NRAO$\,$512. The red curve
represents the obtained respective average power-law fits. The red dots correspond
to the pivot energies listed in Table~\ref{tab:pointlike}.  \label{fig:spectra}}
\end{figure}

The unbinned spectral likelihood analysis was complemented by a \texttt{pointlike} analysis, similar to
the method applied to determine positions and error ellipses for the 1FGL catalog. The approach of 
\texttt{pointlike} is described in Section 4.2 of \citet{2010ApJS..188..405A}. This analysis revealed that a 
significant number of excess $\gamma$-ray photons were detected from the vicinity of NRAO$\,$512 after November 2009 
(15 month). A fit with two point-source models in which source positions were not fixed,
representing contributions by 3C$\,$345 and NRAO$\,$512, yielded significant detections for the 20-month period 
(see Table~\ref{tab:pointlike}). Fig.~\ref{fig:ts_map} shows the radio positions of the three candidate sources together with the EGRET and 1FGL localizations, 
as well as the improved localizations for the two $\gamma$-ray sources found in the 20-month dataset. B3$\,$1640$+$396 is 
19.3\arcmin\ and 16.5\arcmin\ separated from the respective localizations of 3C$\,$345 and NRAO$\,$512. Both $\gamma$-ray counterparts of
 3C$\,$345 and NRAO$\,$512 now coincide with their respective radio positions (within the 95\% error ellipse). This suggests 
that both sources may be $\gamma$-ray emitters. This is firmly established in Section\,\ref{sec:3.3} using the broadband 
temporal characteristics of the sources. There is no evidence for significant $\gamma$-ray emission from the third
candidate source, B3$\,$1640$+$396. The results of \texttt{pointlike} were 
cross-checked with an iterative use of \texttt{gtlike} and \texttt{gtfindsrc}, which confirm the \texttt{pointlike} 
results within errors, see Table~\ref{tab:pointlike}. A 2$\sigma$ upper limit was calculated for B3$\,$1640$+$396 by 
fixing each of the three candidate sources to their radio positions and assuming a $\Gamma=2.3$ spectral shape for 
B3$\,$1640$+$396 and using the previously fitted $\Gamma$ for 3C$\,$345 and NRAO$\,$512. This resulted in a 20-month 
upper limit for the integrated flux of 7.4$\cdot10^{-9}$\,ph\,cm$^{-2}$\,s$^{-1}$. Fig.~\ref{fig:spectra} shows the 
observed binned spectra of the two sources together with their respective power-law fits. The \texttt{pointlike} best-fit 
parameters are listed in Table~\ref{tab:pointlike}. In Fig.~\ref{fig:lightcurve} 
the respective 30-day binned $\gamma$-ray light curves of the two localized sources are shown and are 
discussed below in more detail.

\begin{table}
  \caption{The unbinned spectral likelihood results for the point-source 
localizations and spectra of 3C$\,$345 and NRAO$\,$512. \label{tab:pointlike}}
  \centering
  \begin{tabular}{lll}
  \hline\hline
    & 3C$\,$345 & NRAO$\,$512\\
  \hline
  R.A. (J2000) & 16$^\mathrm{h}$43$^\mathrm{m}$0.24$^\mathrm{s}$ & 16$^\mathrm{h}$40$^\mathrm{m}$44.4$^\mathrm{s}$ \\
  Dec. (J2000) & $+$39\degr48\arcmin22.7\arcsec & $+$39\degr46\arcmin12.0\arcsec \\
  $\Delta$r & 0.35\arcmin & 2.9\arcmin \\
  a (95\%) & 0.0490\degr & 0.0735\degr  \\
  b (95\%) & 0.0442\degr & 0.0626\degr\\
  $\Phi$ & 41.6 & -33.2 \\
  $\Gamma$ (pl) & 2.49$\,\pm\,$0.02 & 2.37$\,\pm\,$0.04\\
  $\Gamma$ (gt) & 2.45\,$\pm$\,0.05 & 2.41\,$\pm$\,0.08\\ 
  Flux (pl) & 1.22$\,\pm\,$0.10 & 0.51$\,\pm\,$0.23\\
  Flux (gt) & 1.13\,$\pm$\,0.13 & 0.67\,$\pm$\,0.13\\
  Pivot Energy & 949 & 1598\\
  TS & 1076 & 246\\
  \hline
  \end{tabular}
  \begin{flushleft}
    \small
    {\bf Notes:}\\
    R.A./Dec. -- Right ascension/declination of the $\gamma$-ray localization;\\ 
    $\Delta r$ -- distance to respective radio source position;\\ 
    a (95\%) -- major axis of 95\% localization error ellipse;\\
    b (95\%) -- minor axis of 95\% localization error ellipse; \\
    $\Phi$ -- Rotation angle of the error ellipse East of North (degrees); \\
    $\Gamma$ -- spectral index of power-law fit, (pl)  are the values obtained from the \texttt{pointlike} fit and (gt) are values from the \texttt{gtlike} fit;\\ 
    Flux -- 20-month average flux (0.1-100\,GeV) determined from the power-law spectral fit 
    in units of 10$^{-7}\,$ph\,cm$^{-2}$\,s$^{-1}$;\\
    Pivot Energy -- in MeV, this is the energy for which there is no correlation 
    between the flux or normalization and spectral index uncertainties; \\
    TS -- Likelihood test statistic value.
  \end{flushleft}
\end{table}

\subsection{20-month of $\gamma$-ray monitoring\label{sec:lc}}

\subsubsection{Long-term variability of 3C\,345 and NRAO\,512\label{sec:long}}

\begin{figure}
  \includegraphics[width=\hsize]{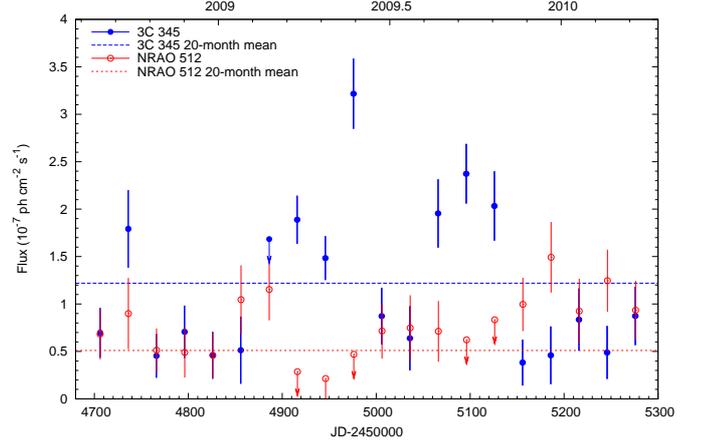}
  \caption{$\gamma$-ray light curves of 3C$\,$345 and NRAO$\,$512 for the first
20~months of \textit{Fermi}-LAT monitoring, obtained through fitting two point-source model components with
power-law spectra to the region of interest placed at the respective counterpart positions. Values were determined from
integrating observations over 30 days within an energy range of
0.1\,$-$\,300\,GeV. Blue filled circles are values obtained for 3C$\,$345, red open circles
are values for NRAO$\,$512. Values with downward arrows represent 2$\sigma$ upper limits shown 
in place of detections with low significance (TS$<$5). The dashed lines plot the respective 20-month
average $\gamma$-ray source flux. Errors are statistical only.\label{fig:lightcurve}}
\end{figure}

To study the long-term variability, light curves were produced for the $\gamma$-ray counterparts associated with 3C$\,$345 and NRAO$\,$512 with 30-day 
binning. The 30-day binned light curve is 
shown in Fig.~\ref{fig:lightcurve}. For most of the data points NRAO$\,$512 did not reach above a TS of 25, except for the last
five months of the 20-month period. This led to the conclusion that NRAO$\,$512 did not produce 
a significant excess of $\gamma$-ray emission over the local background during the first 15 months of LAT operation; 
thus it did not appear in the 1FGL catalog. After 15 months the amount of excess photons from the vicinity of NRAO$\,$512 slowly 
increased, making a detection over the integrated 20-month period possible. The average \textit{Fermi}-LAT flux for the first
15-months of the monitoring period was (0.42\,$\pm$\,0.07)\,$\cdot$\,10$^{-7}$\,ph\,cm$^{-2}$\,s$^{-1}$ compared
to (1.1\,$\pm$\,0.1)\,$\cdot$\,10$^{-7}$\,ph\,cm$^{-2}$\,s$^{-1}$ for the last quarter of the monitoring period (months 15-20).

The observed $\gamma$-ray flux values reveal variability on timescales from
days to months. To quantify this variability, the variability index ($V$) as defined through the $\chi^{2}$ distribution 
\citep[see ][]{2010ApJS..188..405A} was determined. It is computed from the 30-day integrated light curve data shown 
in Fig.~\ref{fig:lightcurve}. Whenever TS\,$<$\,5 the 2$\sigma$ upper limit was computed and its error estimate for that 
interval was replaced with half the difference between that upper limit and its value determined through the unbinned spectral likelihood analysis.
In the absence of variability $V$ is expected to follow a $\chi^2$ distribution with 19 ($=$\,N$_{\mathrm{int}}-1$) degrees 
of freedom. At the 99\% confidence level the light curve is significantly different from a flat one if
$V>36.2$. For 3C\,345 a $V$ of 115.3 was obtained, for NRAO$\,$512 it was 37.7. Both sources are variable according to the 99\% 
confidence interval with higher variability observed from 3C$\,$345. 

In the shorter binned light curve of NRAO$\,$512, $\sim$63\% of the derived values had a TS\,$<$\,5 (i.e. non-detections) compared to 
$\sim$48\% for 3C$\,$345. Because NRAO$\,$512 did not show a significant flux for most of the 
time on integrations of five days or less and with the observation that the flux of 3C\,345 dominates, the model component 
of NRAO\,512 was removed and light curves were produced with a single 
component fixed at the position of the $\gamma$-ray localization of 3C$\,$345 listed in Table~\ref{tab:pointlike}. This
reduced the amount of noise in the light curve of 3C$\,$345 and the number of values with a TS\,$<$\,5 was only 14\%.

\subsubsection{Short-term variability of 3C\,345\label{sec:short}}

\begin{figure}
  \includegraphics[width=\hsize]{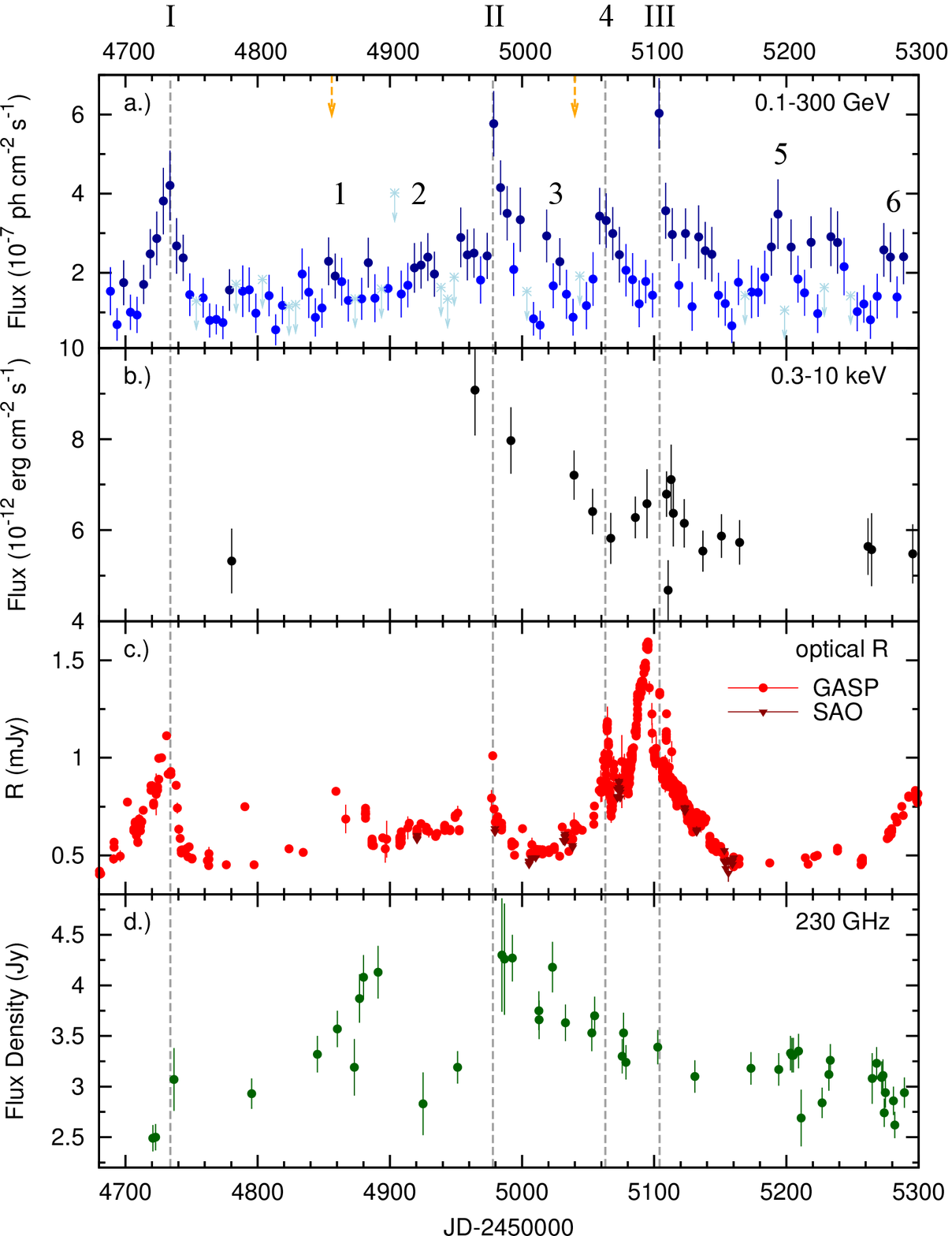}
  \caption{Multiwavelength light curves of 3C$\,$345 for the 20-month period discussed in this paper, 
from top to bottom for: a) $\gamma$ rays observed by \textit{Fermi} LAT between 0.1 and 300\,GeV, values (filled circles) 
are five-day integrated, 2$\sigma$ upper limits for five-day integrations are shown as light-blue crosses with downward arrows where TS\,$<$\,5. 
Dark-blue filled circles have a TS\,$>$\,25, the others are in the range 5\,$<$\,TS\,$\leq$\,25. In contrast
to Fig.~\ref{fig:lightcurve}, the $\gamma$-ray emission was fitted by a single power-law component placed
at the position of 3C\,345. b) X-rays observed by \textit{Swift} XRT between 0.3 and 10\,keV. c) optical (R-band) 
observations performed by GASP and SAO. d) millimeter radio observations by the SMA at 230\,GHz.
The two orange downward arrows on top indicate the observation epoch for which a new feature was detected in the pc-scale radio jet of 3C\,345, 
see Section~\ref{sec:3.4} and \citet{SchinzelFmJ2010}. Three dominant flare events of high significance observed on two-day time scales
were labeled with I, II, and III. Several weaker events were identified and labeled with numbers from 1 -- 6.
\label{fig:mwlc} }

\end{figure}

For the remainder of this paper the discussion is based on light 
curves produced with a single point-source model component. The five-day 
binned light curve of 3C$\,$345 is shown in Fig.~\ref{fig:mwlc} for comparison
with observations at other wavelengths.   

To investigate shorter term variability in the observed $\gamma$-ray emission
of 3C\,345, several prominent light-curve ``flare events'' were
identified and parameters such as rise time, fall time and
time of the peak were extracted using the five-day and two-day integrated
light curves. These parameters can be used to constrain the size of emission
regions and cooling-times. Table~\ref{tab:gammaevents} summarizes these
results, listing characteristic parameters for all $\gamma$-ray flare events for which the peak
had a significance of TS\,$>$\,25 in the two-day light curve. This corresponded to a flux greater than 
$3.0\,\cdot$\,10$^{-7}$\,ph\,cm$^{-2}$\,s$^{-1}$. The time of onset 
and end of an individual flare event was determined through a change in the sign 
of the gradient at the rising and declining slope of the light curve around the peak of a flare event.

\begin{table*}
  \caption{Characteristics of prominent $\gamma$-ray events during
the first 20 months of the \textit{Fermi}-LAT observations. \label{tab:gammaevents}}
  \centering
  \begin{tabular}{c c c c c c c c}
  \hline\hline
  $t_\mathrm{peak}$ & $t_\mathrm{start}$ & $t_\mathrm{stop}$ &
$\Delta t$ & $\Delta t_\mathrm{rise}$ & $\Delta t_\mathrm{fall}$ &
$S_\gamma^\mathrm{max}$ & (*)\\
JD($\dagger$) & JD($\dagger$) & JD($\dagger$) & days & days & days &
$10^{-7}\,\frac{\mathrm{ph}}{\mathrm{cm}^{2}\,\mathrm{s}}$\\
  \hline
    4734 & 4709 & 4754 & 45 & 25 & 20 & $5.5\pm1.5$
& $\mathrm{I}$ \\
    4856 & 4844 & 4874 & 30 & 12 & 18 & $3.6\pm1.1$ &
$\mathrm{1}$ \\
    4916 & 4904 & 4944 & 40 & 12 & 28 & $4.0\pm1.2$ &
$\mathrm{2}$ \\
    4978/4998 & 4969 & 5014 & 45 & 9 & 16 & $5.7\pm1.3$/$5.4\pm1.5$ &
$\mathrm{II}$ \\
    5020 & 5014 & 5044 & 30 & 6 & 24 & $4.5\pm1.2$ &
$\mathrm{3}$ \\
    5056/5070 & 5044 & 5089 & 45 & 12 & 19 & $4.4\pm1.4$/$4.5\pm1.2$ &
$\mathrm{4}$ \\
    5104 & 5099 & 5119 & 20 & 5 & 15 & $9.1\pm1.6$ &
$\mathrm{III}$\\
    5194 & 5169 & 5214 & 45 & 25 & 20 & $4.7\pm1.4$ &
$\mathrm{5}$\\
    5278 & 5264 &  & & 14 & & $3.6\pm1.1$ &
$\mathrm{6}$\\
  \hline
  Average values: & & &  38 & 13 & 20 & 5.0\,$\pm$\,0.5\\
  \hline
  \end{tabular}
  \begin{flushleft}
  {\bf Notes:} 
$t_\mathrm{peak}$ -- in units of JD($\dagger$)\,$=$\,JD-2450000, peak value was obtained from
the two-day integrated light curve, all values have an error of $\pm$\,1 day;\\
$t_\mathrm{start}$, $t_\mathrm{stop}$ -- in
units of JD($\dagger$)\,$=$\,JD-2450000, all values have an error of $\pm$\,3
days;\\
  $\Delta t$, $\Delta t_\mathrm{rise}$, $\Delta t_\mathrm{fall}$
-- in units of days, all values with error of $\pm$\,4 days;\\ $S_\gamma^\mathrm{max}$ -- two-day integrated flux value of 
the $\gamma$-ray event peak value;\\
(*) Labels are split into two sub-categories according to their peak flux and
significance of detection, denoted by Roman and Arab numbers, see
Fig.~\ref{fig:mwlc} for details.\\

  \end{flushleft}
\end{table*}

Three dominant flare events of at least $5.0\cdot10^{-7}\,$ph\,cm$^{-2}$\,s$^{-1}$ and a
significance of above 7$\sigma$ were observed on two-day
time scales, one during 2008 and two during
2009. They were labeled with $\mathrm{I}$, $\mathrm{II}$ and $\mathrm{III}$ in Fig.~\ref{fig:mwlc}. 
Several weaker flares were identified and labeled with numbers from 1 to 6. On average, the flares 
had a duration of 38 days a slightly faster rise than fall time and an average peak flux of
$(5.0\pm0.5)\cdot10^{-7}\,$ph\,cm$^{-2}$\,s$^{-1}$ (two-day integrations centered on the peak).

\subsection{Identification of $\gamma$-ray counterparts\label{sec:3.3}}

Following the $\gamma$-ray outburst on October 2nd, 2009, \textit{Swift}
observations were triggered on the candidate sources 3C$\,$345, NRAO$\,$512 and
B3$\,$1640$+$396. X-ray spectra and optical magnitudes were obtained from these observations.
A summary of relevant measurements is listed in
Tables~\ref{tab:XRTdata} and \ref{tab:opticaldata}.

The brightest source of the three candidates in X-ray as well as in optical is 3C$\,$345.
On January 26th, 2009, 3C$\,$345 and B3$\,$1640$+$396 were detected in
the optical R-band at respective magnitudes of 16.48\,$\pm$\,0.02 and
18.6\,$\pm$\,0.1. On October 8th, 2009, observations of 3C$\,$345 and NRAO$\,$512 showed
respective optical U-band magnitudes of 16.45\,$\pm$\,0.08 and
18.64\,$\pm$\,0.04. Both NRAO$\,$512 and B3$\,$1640$+$396 were a factor of 7-8 fainter
than 3C$\,$345 at optical wavelengths. At X-ray energies, as seen from
Table~\ref{tab:XRTdata}, NRAO$\,$512 and B3$\,$1640$+$396 had similar
fluxes, which were a factor of 8-9 fainter than 3C$\,$345. Archival
\textit{Swift}-UVOT and XRT data from observations of NRAO$\,$512 made in January
2007 have an average optical U magnitude of 18.0\,$\pm$\,0.1 and an X-ray flux of
(8.3\,$\pm$\,2.3)$\cdot$10$^{-13}$\,erg\,cm$^{-2}$\,s$^{-1}$, which are
consistent with the recently observed values. Quasi-simultaneous mm observations 
between October and December 2009 determined the fluxes of 3C$\,$345, NRAO$\,$512 and B3$\,$1640$+$396 
to be $\sim$\,3.0, $\sim$\,0.4 and $\sim$\,0.1$-$0.3\,Jy. The respective flux ratios are consistent with the ones in the
optical/UV and X-rays. The mm radio flux density of 3C$\,$345 was at least a
factor of 7 higher than for NRAO$\,$512 and B3$\,$1640+396.

No significant excess of $\gamma$-ray emission from NRAO$\,$512 was observed during the
first 15-months of the \textit{Fermi}-LAT operation. At the end of 2009, NRAO$\,$512 showed an increasing
trend in its 15\,GHz VLBI radio flux observed by the MOJAVE team
\citep{2009AJ....137.3718L}\footnote{\href{http://www.physics.purdue.edu/MOJAVE/sourcepages/1638+398.shtml}{www.physics.purdue.edu/MOJAVE/sourcepages/1638+398.shtml}}.
Indeed, the radio flux reached a low activity level around May 2009 and more than doubled until about July 2010 with the mm-radio
flux density reaching a peak value of over 0.6\,Jy in August 2010.
A similar trend was observed in optical emission. Within one year, the optical flux increased by a factor
of $\sim$2.2 (see Table~\ref{tab:opticaldata}). In X-rays a flux increase of at least 20-30\% was observed, 
but because of the low statistics (less than 200 photons) it is difficult to compare the observations. An increased number of 
$\gamma$-ray photons was detected spatially consistent with NRAO$\,$512 during the last quarter of the 20-month period.
The excess is not high enough to affect the short-term bins of the light curve of 3C$\,$345 significantly, however, as 
was shown in Section~\ref{sec:3.1}, the integrated 20-month dataset made it possible to detect a second 
point source west of 1FGL\,J1642.5+3947, which is consistent with
the radio position of NRAO$\,$512 (see Fig.~\ref{fig:ts_map}). 
Altogether this identifies NRAO$\,$512 as a $\gamma$-ray source.

\subsection{Multi-wavelength analysis of $\gamma$-ray emission from 3C\,345\label{sec:3.4}}

During 2009 nine optical observatories, listed in Section~\ref{sec:optical},
regularly monitored
3C$\,$345, mainly as part of the GASP-WEBT program. This provided a densely
sampled optical R-band light curve (Fig.~\ref{fig:mwlc}). Visual
comparison of the $\gamma$-ray and optical light curves indicates a
likely correspondence between at least four $\gamma$-ray/optical
flare events: Flare $\mathrm{I}$, $\mathrm{II}$, 4 and $\mathrm{III}$. With the
addition of X-ray and mm-radio monitoring data,   
the following aspects are of particular interest:
\begin{enumerate}
 \item A possible fast optical brightening of at least 0.3 mag with a duration of only a few days 
(Fig.~\ref{fig:mwlc}) was observed (JD\,2454978), coincident with the peak of Flare II. Unfortunately, 
the optical monitoring had a gap between JD\,2454953 and JD\,2454977. At the 
same time the mm radio light curve showed a maximum of around 4.3\,Jy. Moreover, a high X-ray flux 
was observed in the same period of the Flare II, followed by a decrease similar to what was observed in $\gamma$ 
and radio bands. Even if the maximum X-ray flux was observed two weeks before Flare II, the $\gamma$-ray event 
lacks strictly simultaneous X-ray observations, thus we cannot exclude that a simultaneous X-ray flare was missed.
 \item The broad optical flaring episode observed around September/October 2009 had a total
duration of about five months, with shorter timescale sub-structures that
correspond to activity variations observed also at $\gamma$- and X-rays. 
The peak in the optical was observed on September 19th, 2009 (JD\,2455094), and had no
similarity in profile with events at other wavelengths (see Fig.~\ref{fig:mwlc}).
 \item Ten days after the peak of the optical light curve, the strong $\gamma$-ray
Flare $\mathrm{III}$ was observed. The optical peak was followed by fast optical
variability at a significant amplitude of 0.3$-$0.5~mag, with an elevated
magnitude observed on September 29th on the same day as $\gamma$-ray Flare
$\mathrm{III}$.
 \item Flares 1 and 3 can be related to the appearance of new features in the pc-scale radio jet of 3C$\,$345 at
43\,GHz (see \citealt{SchinzelFmJ2010}). 
 \item All but Flares 5 and 6 occur before the significant excess of $\gamma$-ray emission was
observed near NRAO$\,$512. However, for at least Flare 6, a correspondence in optical 
variability was found for 3C$\,$345; unfortunately, a lack of multiwavelength data around Flare 5
makes it impossible to identify it with either 3C$\,$345 or NRAO$\,$512.
\end{enumerate}

\begin{figure}
  \includegraphics[width=\hsize]{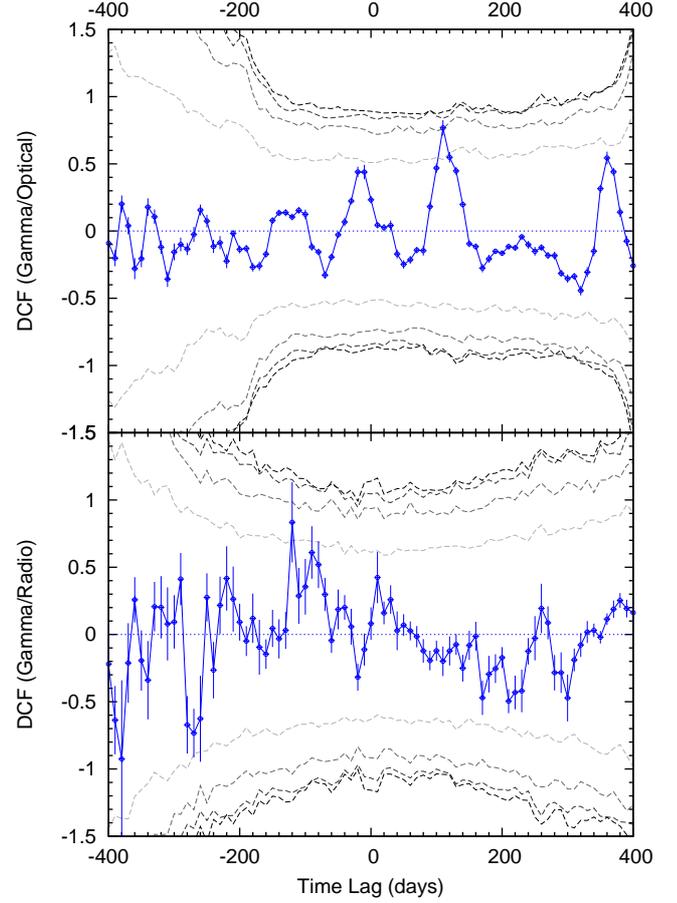}
  \caption{Correlation between the five-day integrated $\gamma$-ray light curve and
the optical (top) and radio (bottom) light curves of 3C\,345. The
correlation is obtained by calculating the discrete correlation function
(DCF) with ten-day binning. A positive time-lag corresponds to the
$\gamma$-ray variations preceding the variations in the other two
bands. The significance of the correlation is illustrated by 3$\sigma$ (99.7\%)
significance contours calculated from simulated, uncorrelated data
with different slopes, $\beta$, of the power spectral density
(PSD). The contours are drawn for the PSD slopes $\beta = $ 1.0, 1.5, 2.0, and
2.5 (indicated by progression of colors, from light gray to black).
For the simulated data, values of the correlation coefficient
exceeding $\pm 1$ are an inherent feature of the method, resulting from
differences in overlap of the time series at different time lags \citep[cf.][and the description
of the applied method in Section~\ref{sec:3.4}]{1988ApJ...333..646E, 1995A&A...300..707T}.
\label{fig:dcf}}
\end{figure}

To quantify the cross-correlation between light curves, a discrete correlation was calculated
\cite[DCF;][]{1988ApJ...333..646E} using the 20-month \textit{Fermi}-LAT light curve and the
optical R-band light curve between JD~2454680 (August 1st, 2008) and 2455299
(April 12th, 2010), as well as the radio SMA light curve between JD~2454721 (September
11th, 2008) and 2455289 (April 2nd, 2010). Fig.~\ref{fig:dcf} (top) plots the resulting DCF 
comparing optical and $\gamma$-ray light curves using a five-day $\gamma$-ray integration and a ten-day timelag binning. 
This yields a number of distinct correlation peaks, of which the two most significant are discussed. In the first peak the 
optical leads the $\gamma$-ray flux by (15\,$\pm$\,10)\,days with a correlation
coefficient of 0.44\,$\pm$\,0.05. The second peak corresponds to the optical flux lagging the $\gamma$-rays by
110\,$\pm$\,5\,days with a correlation coefficient of 0.77\,$\pm$\,0.06.
To test the robustness of these peaks in the DCF, we artificially removed the 
points corresponding to the peak $\gamma$-ray flux from Flares II and III and recalculated the DCF.
The same behavior was observed, with the $\gamma$-ray flux leading the optical by 20 days and lagging by 110 days.
This shows that the results obtained do not solely depend on these two events. Finally, the 
DCF between the $\gamma$-ray and the SMA light curve of 3C\,345 was calculated, which yielded
a noisy distribution of correlation peaks (also see Fig.~\ref{fig:dcf}). The strongest peak indicates 
that the radio emission leads the $\gamma$ rays by 120\,$\pm$\,5\,days with a correlation coefficient of 0.83\,$\pm$\,0.30.

The statistical significance of the cross-correlation peaks was investigated
using Monte Carlo simulations, following \citet{2008ApJ...689...79C} 
and \citet{FmJ2010} and assuming that the noise
properties of the light curves can be described by a power-law power spectral density (PSD; $\propto 1/f^{-\beta}$) 
with an exponent that depends on the particular energy band of the emission. 

In order to estimate the significance of the $\gamma$-ray/optical cross correlation,
an attempt was made to calculate and characterize the PSD shape from the 
presented data directly, despite the limited length (20 months) and sampling ($>$ 1\,day) of the light curves.
For the optical and $\gamma$-ray data, the light curves available support a range 
of time scales corresponding to about one decade in the frequency domain. The
respective range is even smaller for the more sparsely sampled radio data.
The PSDs were calculated following Section 3.1 in \citet{2002MNRAS.332..231U}. 
The resulting power-law slopes, $\beta$, are in the range of 1.2--0.7. Values
of $\beta \gg 2.0$ are not consistent with the calculated PSD shapes and can be ruled out. 
However, \citet{2002MNRAS.332..231U} demonstrated that aliasing effects add to the power spectral 
slopes and could potentially flatten the observed power spectrum.
The relatively high noise and short time span of the data makes this range of $\beta$ rather unreliable. 
Monte Carlo simulations for these values of $\beta$ yield extremely high cross-correlation 
significances of $>$99.999\% for the $\gamma$-ray/optical light curves and $>$99.98\% for the $\gamma$-ray/radio case. 
It should be noted that a reliable determination of the PSD from observational data is usually done with a much larger
dynamic range (e.g. covering 4--5 orders of magnitude in the frequency scale), which is also the case in \citet{2002MNRAS.332..231U}. 
The radio, optical and $\gamma$-ray data discussed in our paper do not provide 
such an extended coverage in the frequency space. We adopt therefore a more cautious approach, exploring 
a range of power-law indices based on published values.

In the $\gamma$-ray regime $\beta_{\gamma} = 1.7\,\pm\,0.3$ for BL
Lacs and  $\beta_{\gamma} = 1.4\,\pm\,0.1$ for flat spectrum radio quasars \citep{2010ApJ...722..520A}. 
Instead, for 3C\,279, it has been found that $\beta_{radio} = 2.3\,\pm\,0.5$ 
at 14.5 GHz and $\beta_{optical} = 1.7\,\pm\,0.3$ in the R-band
\citep{2008ApJ...689...79C}. More recently \citet{2011arXiv1101.3815C} found an 
average optical power-law index of $\beta_{optical} =
1.6\,\pm\,0.3$ for a sample of six blazars in the R-band. 
Simulated light curves with PSD power-law indices in these ranges can provide 
reliable estimates on the significance of the cross-correlations.
The simulations use light curves of the same length as the data and simulated up 
to a maximum frequency of 1 day$^{-1}$ using the method described by \citet{1995A&A...300..707T}.
Each case uses 10000 pairs of light curves, which are sampled and
cross-correlated in the same fashion as the observed light curves. From these simulated data the distribution of
correlation coefficients for each time bin can be obtained for the case of non
correlated time series. Fig.~\ref{fig:dcf} presents the $3\sigma$ (99.7\%) 
significance levels for all tested PSD shape combinations.

The $\gamma$-ray/optical correlation has a significance of higher than 3$\sigma$ (99.7\%) for the peak at a timelag of 110 days
in most of the tested cases. It
is the most significant peak in the DCF and it has a moderately high confidence level ($> 98.5\%$) even for the least
significant case ($\beta_{\gamma} = 2.5$ and $\beta_{\mathrm{optical}} = 2.5$; highest dashed line in the top panel of 
Fig.~\ref{fig:dcf}) among the $\beta$ values considered. 
The $\gamma$-ray/optical peak at a timelag of about $-$20\,days has a much lower confidence level ($> 67.6\%$) in the least significant 
case and it remains below the 3$\sigma$ level at all times. The
$\gamma$-ray/radio cross-correlation at timelag $-$120 days has a confidence level of $> 90.7 \%$ in
the least significant case ($\beta_{\gamma} = 2.5$ and $\beta_{\mathrm{radio}} = 3.0$).

The source 3C$\,$345 is by far the brightest of the three quasars in the $\gamma$-ray region studied
during the presented time period. In all bands it was consistently 
brighter than NRAO$\,$512 and B3$\,$1640$+$396. The improved \textit{Fermi} position combined with 
multi-wavelength activity and a moderate significance of correlated optical/$\gamma$-ray variability,
using a plausible range of $\beta$ in the $\gamma$-ray and optical PSDs,
strongly suggest an identification of 3C\,345 as $\gamma$-ray source and the main contributor
to the observed $\gamma$-ray flux in 1FGL\,J1642.5+3947.

\section{Discussion and summary}

Properties of the $\gamma$-ray emitting region reported in the first three-month
Bright Source List \citep{2009ApJS..183...46A} as well as in the first 11-month
\textit{Fermi}-LAT source catalog \citep{2010ApJS..188..405A,
2010ApJ...715..429A} were compared to the 20-month dataset presented here.
Localizations of the sources in this region were improved over
the previously reported values, leading to the detection of two $\gamma$-ray point sources, one consistent 
with 1FGL\,J1642.5+3947 and a new $\gamma$-ray source $\sim$\,0.4\degr\ west of it. 3C$\,$345 and NRAO$\,$512 were
identified as the counterparts for these $\gamma$-ray sources. Their $\gamma$-ray 
spectra have a spectral slope of $\Gamma=$\,2.4--2.5 typical for that of flat spectrum radio quasars \citep{2010ApJ...715..429A}.
No significant spectral break was observed in the spectrum of either source, unlike in the case of 3C 454.3 \citep{2010ApJ...721.1383A}, 
however, the statistics at the highest energies are poor, and the presence of a spectral break at E$>$10\,GeV is not ruled out.

The $\gamma$-ray emission from 3C$\,$345 was identified based on an improved $\gamma$-ray counterpart
localization and multi-wavelength activity including correlations of moderate significance found
between the optical and radio variability, using a plausible range of $\beta$ in the $\gamma$-ray and optical PSDs, 
and major $\gamma$-ray events observed
by \textit{Fermi} LAT during its first 20 months of operation.
EGRET observed 3C$\,$345
between 1991 and 1994 during a time of enhanced radio activity. An upper limit of 
2.5\,$\cdot$\,10$^{-7}$\,ph\,cm$^{-2}$\,s$^{-1}$ for the viewing period September 12th - 19th, 1991, was reported
\citep{1994ApJS...94..551F}. However, \citet{2008A&A...489..849C} report the detection of a $\gamma$-ray source
that could be consistent with 3C\,345 in the period April 23rd - May 7th, 1996 with a $\gamma$-ray flux of 
(3.5\,$\pm$\,0.8)\,$\cdot$\,10$^{-7}$\,ph\,cm$^{-2}$\,s$^{-1}$, but it was noted that the association
with 3C$\,$345 remains unclear because of a possible confusion with Mrk\,501. In the \textit{Fermi}-era, 
with an increased sensitivity and
the possibility of long-term integrations, 3C$\,$345 was already detected during the
first three months of operation at a level of $\sim$\,18$\sigma$ and after 20 months
it reached over 30$\sigma$. However, owing to the unclear association of 3C\,345 with EGR\,J1642+3940,
the three-month data were falsely associated with CLASS\,J1641+3035 (B3\,1640+396), which, as shown in this paper,
did not produce a significant excess of $\gamma$-ray emission above the detection threshold of \textit{Fermi}-LAT.
Thus the simplest explanation for the previous non-detection of 3C$\,$345 is that its emission was just below or 
at the detection limit of EGRET.

These findings end the decade-long
debate whether 3C$\,$345 is $\gamma$-ray loud \citep{1997ApJ...480..596U}.
The identification of 1FGL\,J1642.5+3947 with 3C$\,$345 is further supported by
the connection of radio-emission of the parsec-scale jet with the observed
$\gamma$-ray emission \citep{SchinzelFmJ2010}. Here not only individual flares were correlated with radio
events, but an unambigious long-term trend in $\gamma$-ray emission was observed, which matched the trend 
of the radio flux density light curve of the inner jet at 43\,GHz. The trend
and correlated radio events relate to a distance of $\sim$10$\,$pc in the source
frame and questions existing scenarios that place
the $\gamma$-ray emission at sub-pc distances from the central engine \citep[e.g.,][]{2010MNRAS.405L..94T}.
This supports the arguments given by \citet{1997ApJ...480..596U} that the observed $\gamma$-ray emission
is related to the relativistic jet on pc scales. 

The cross-correlation found between $\gamma$-ray and optical light curves at 110\,days had a
moderate confidence level of $>$\,98.5\%, whereas the lag at -20\,days had a confidence level of only
$>$\,67.6\% using a plausible range of $\beta$ in the $\gamma$-ray and optical PSDs. If typical values for the power spectral density (PSD) shapes are assumed 
($\beta_{\gamma} = 1.5$ and $\beta_{\mathrm{optical}} = 1.5$; second to lowest dashed line in the top panel of 
Fig.~\ref{fig:dcf}), the confidence level 
remains low at $>$\,89.3\%. The weak correlation found between the $\gamma$-ray and 
230\,GHz radio light curves had a worst-case confidence level of $>$\,89.3\%. Assuming a possibly more 
typical combination of PSD shapes ($\beta_{\gamma} = 1.5$ and $\beta_{\mathrm{radio}} = 2.0$), 
a confidence level of $>$\,96.1\% is obtained. If this weak correlation holds, it  
could suggest a connection between the radio emission and production of $\gamma$-rays. 
A timelag of $\sim$-120 days suggests the mm-radio emission precedes the $\gamma$-rays. \citet{2010ApJ...722L...7P} showed 
for a larger sample of radio-loud quasars 
that on average the 15\,GHz radio emission lags $\sim$1.2 months behind $\gamma$-rays. The result presented here   
indicates that between 15\,GHz and 230\,GHz it is possible to directly observe the $\gamma$-ray emitting region, 
as already implied by \citet{1997ApJ...480..596U}, which supports the argument of $\gamma$-ray production in the relativistic 
jet on pc scales. However, in contrast to these, a more confident cross-correlation result of $\gamma$-rays leading by $\sim$80 and
 $\sim$50\,days was recently obtained for two prominent BL Lac type objects, OJ\,287 \citep{2011ApJ...726L..13A}
and AO\,0235+164 \citep{2011arXiv1105.0549A}, respectively. Doubts on the indicated
radio/$\gamma$-ray correlation of 3C\,345 remain in this problem and should be investigated 
more thoroughly with a longer and denser sampled lightcurve 
at mm wavelengths.

The $\gamma$-ray emission from NRAO$\,$512, a high-redshift quasar, was identified based 
on a correlated increase in multiwavelength emission from radio up to GeV energies between 2009 and 2010. 
At monthly timescales the source started to be significant around November 2009. 
This increase of $\gamma$-ray flux led to the first detection of the now identified new 
$\gamma$-ray counterpart west of 3C$\,$345. The LAT flux quoted in this paper for the first 15 months was below
the EGRET detection threshold, offering a plausible explanation of why EGRET did not see emission from NRAO$\,$512,
except possibly during a $\gamma$-ray flare event.

The radio, optical and X-ray flux ratios between 3C$\,$345, NRAO$\,$512
and B3$\,$1640$+$396 could be used to gauge their individual $\gamma$-ray photon 
contributions and might be able to provide an indication of false associations for $\gamma$-ray
photons from this part of the sky. Multiwavelength monitoring, in addition to the continuous all-sky monitoring by
\textit{Fermi} LAT, will help to improve the localization of $\gamma$-ray emission from this region and is necessary
for a deeper variability study for any of these sources in the future. Ultimately this might lead
to the detection of $\gamma$-ray emission from B3$\,$1640$+$396 as well.

\begin{acknowledgements}
The authors would like to thank Ann Wehrle, Benoit Lott, Seth Digel, Gino
Tosti, Dave Thompson, and Greg Taylor for their helpful comments and useful discussions,
which improved the quality of this manuscript. We thank the A\&A referee for 
extensive comments.

The \textit{Fermi} LAT Collaboration acknowledges generous ongoing support
from a number of agencies and institutes that have supported both the
development and the operation of the LAT as well as scientific data analysis.
These include the National Aeronautics and Space Administration and the
Department of Energy in the United States, the Commissariat \`a l'Energie
Atomique and the Centre National de la Recherche Scientifique / Institut
National de Physique Nucl\'eaire et de Physique des Particules in France, the
Agenzia Spaziale Italiana and the Istituto Nazionale di Fisica Nucleare in
Italy, the Ministry of Education, Culture, Sports, Science and Technology
(MEXT), High Energy Accelerator Research Organization (KEK) and Japan Aerospace
Exploration Agency (JAXA) in Japan, and the K.~A.~Wallenberg Foundation, the
Swedish Research Council and the Swedish National Space Board in Sweden.
Additional support for science analysis during the operations phase is gratefully 
acknowledged from the Istituto Nazionale di Astrofisica in Italy and the Centre National d'Etudes Spatiales in France.

Frank Schinzel and Kirill Sokolovsky were supported for this
research through a stipend from the International Max Planck Research School
(IMPRS) for Astronomy and Astrophysics at the Universities of Bonn and Cologne.

M. Villata organized the optical-to-radio observations by GASP-WEBT as the president of the collaboration.
Additional support for science analysis during the operations phase is
gratefully acknowledged from the Istituto Nazionale di Astrofisica in Italy and
the Centre National d'\'Etudes Spatiales in France. The St.Petersburg University 
team acknowledges support from Russian RFBR foundation via grant 09-02-00092.
Acquisition of the MAPCAT data is supported in part by the Spanish Ministry of
Science and Innovation and the Regional Government of Andaluc\'{i}a through
grants AYA2007-67626-C03-03 and P09-FQM-4784, respectively. The Abastumani 
team acknowledges financial support by the Georgian National Science Foundation 
through grant GNSF/ST08/4-404. A. Marscher and S. Jorstad received support for
this work from NASA grants NNX08AV65G, NNX08AV61G, and NNX09AT99G, and National
Science Foundation grant AST-0907893 to Boston University.

The Submillimeter Array is a joint project between the Smithsonian Astrophysical
Observatory and the Academia Sinica Institute of Astronomy and Astrophysics and
is funded by the Smithsonian Institution and the Academia Sinica. This paper is 
partly based on observations carried out at the German-Spanish
Calar Alto Observatory, which is jointly operated by the MPIA and the IAA-CSIC.
This research has made use of NASA's Astrophysics Data System, the SIMBAD database,
operated at CDS, Strasbourg, France and the NASA/IPAC Extragalactic Database (NED) 
which is operated by the JPL, Caltech, under contract with NASA. This research has 
made use of data from the MOJAVE database that is maintained by the MOJAVE team 
\citep{2009AJ....137.3718L}.

\end{acknowledgements}

\bibliographystyle{aa}
\bibliography{references}

\begin{thebibliography}{59}
\expandafter\ifx\csname natexlab\endcsname\relax\def\natexlab#1{#1}\fi

\bibitem[{{Abazajian} {et~al.}(2009){Abazajian}, {Adelman-McCarthy},
  {Ag{\"u}eros}, {Allam}, {Allende Prieto}, {An}, {Anderson}, {Anderson},
  {Annis}, {Bahcall}, {Bailer-Jones}, {Barentine}, {Bassett}, {Becker},
  {Beers}, {Bell}, {Belokurov}, {Berlind}, {Berman}, {Bernardi}, {Bickerton},
  {Bizyaev}, {Blakeslee}, {Blanton}, {Bochanski}, {Boroski}, {Brewington},
  {Brinchmann}, {Brinkmann}, {Brunner}, {Budav{\'a}ri}, {Carey}, {Carliles},
  {Carr}, {Castander}, {Cinabro}, {Connolly}, {Csabai}, {Cunha}, {Czarapata},
  {Davenport}, {de Haas}, {Dilday}, {Doi}, {Eisenstein}, {Evans}, {Evans},
  {Fan}, {Friedman}, {Frieman}, {Fukugita}, {G{\"a}nsicke}, {Gates},
  {Gillespie}, {Gilmore}, {Gonzalez}, {Gonzalez}, {Grebel}, {Gunn},
  {Gy{\"o}ry}, {Hall}, {Harding}, {Harris}, {Harvanek}, {Hawley}, {Hayes},
  {Heckman}, {Hendry}, {Hennessy}, {Hindsley}, {Hoblitt}, {Hogan}, {Hogg},
  {Holtzman}, {Hyde}, {Ichikawa}, {Ichikawa}, {Im}, {Ivezi{\'c}}, {Jester},
  {Jiang}, {Johnson}, {Jorgensen}, {Juri{\'c}}, {Kent}, {Kessler}, {Kleinman},
  {Knapp}, {Konishi}, {Kron}, {Krzesinski}, {Kuropatkin}, {Lampeitl},
  {Lebedeva}, {Lee}, {Lee}, {Leger}, {L{\'e}pine}, {Li}, {Lima}, {Lin}, {Long},
  {Loomis}, {Loveday}, {Lupton}, {Magnier}, {Malanushenko}, {Malanushenko},
  {Mandelbaum}, {Margon}, {Marriner}, {Mart{\'{\i}}nez-Delgado}, {Matsubara},
  {McGehee}, {McKay}, {Meiksin}, {Morrison}, {Mullally}, {Munn}, {Murphy},
  {Nash}, {Nebot}, {Neilsen}, {Newberg}, {Newman}, {Nichol}, {Nicinski},
  {Nieto-Santisteban}, {Nitta}, {Okamura}, {Oravetz}, {Ostriker}, {Owen},
  {Padmanabhan}, {Pan}, {Park}, {Pauls}, {Peoples}, {Percival}, {Pier}, {Pope},
  {Pourbaix}, {Price}, {Purger}, {Quinn}, {Raddick}, {Fiorentin}, {Richards},
  {Richmond}, {Riess}, {Rix}, {Rockosi}, {Sako}, {Schlegel}, {Schneider},
  {Scholz}, {Schreiber}, {Schwope}, {Seljak}, {Sesar}, {Sheldon}, {Shimasaku},
  {Sibley}, {Simmons}, {Sivarani}, {Smith}, {Smith}, {Smol{\v c}i{\'c}},
  {Snedden}, {Stebbins}, {Steinmetz}, {Stoughton}, {Strauss}, {Subba Rao},
  {Suto}, {Szalay}, {Szapudi}, {Szkody}, {Tanaka}, {Tegmark}, {Teodoro},
  {Thakar}, {Tremonti}, {Tucker}, {Uomoto}, {Vanden Berk}, {Vandenberg},
  {Vidrih}, {Vogeley}, {Voges}, {Vogt}, {Wadadekar}, {Watters}, {Weinberg},
  {West}, {White}, {Wilhite}, {Wonders}, {Yanny}, {Yocum}, {York}, {Zehavi},
  {Zibetti}, \& {Zucker}}]{2009ApJS..182..543A}
{Abazajian}, K.~N., {Adelman-McCarthy}, J.~K., {Ag{\"u}eros}, M.~A., {et~al.}
  2009, \apjs, 182, 543

\bibitem[{{Abdo} {et~al.}(2010{\natexlab{a}}){Abdo}, {Ackermann}, {Ajello},
  {Allafort}, {Antolini}, {Atwood}, {Axelsson}, {Baldini}, {Ballet},
  {Barbiellini}, {Bastieri}, {Baughman}, {Bechtol}, {Bellazzini}, {Belli},
  {Berenji}, {Bisello}, {Blandford}, {Bloom}, {Bonamente}, {Bonnell},
  {Borgland}, {Bouvier}, {Bregeon}, {Brez}, {Brigida}, {Bruel}, {Burnett},
  {Busetto}, {Buson}, {Caliandro}, {Cameron}, {Campana}, {Canadas}, {Caraveo},
  {Carrigan}, {Casandjian}, {Cavazzuti}, {Ceccanti}, {Cecchi}, {{\c C}elik},
  {Charles}, {Chekhtman}, {Cheung}, {Chiang}, {Cillis}, {Ciprini}, {Claus},
  {Cohen-Tanugi}, {Conrad}, {Corbet}, {Davis}, {DeKlotz}, {den Hartog},
  {Dermer}, {de Angelis}, {de Luca}, {de Palma}, {Digel}, {Dormody}, {Silva},
  {Drell}, {Dubois}, {Dumora}, {Fabiani}, {Farnier}, {Favuzzi}, {Fegan},
  {Ferrara}, {Focke}, {Fortin}, {Frailis}, {Fukazawa}, {Funk}, {Fusco},
  {Gargano}, {Gasparrini}, {Gehrels}, {Germani}, {Giavitto}, {Giebels},
  {Giglietto}, {Giommi}, {Giordano}, {Giroletti}, {Glanzman}, {Godfrey},
  {Grenier}, {Grondin}, {Grove}, {Guillemot}, {Guiriec}, {Gustafsson},
  {Hadasch}, {Hanabata}, {Harding}, {Hayashida}, {Hays}, {Healey}, {Hill},
  {Horan}, {Hughes}, {Iafrate}, {J{\'o}hannesson}, {Johnson}, {Johnson},
  {Johnson}, {Johnson}, {Kamae}, {Katagiri}, {Kataoka}, {Kawai}, {Kerr},
  {Kn{\"o}dlseder}, {Kocevski}, {Kuss}, {Lande}, {Landriu}, {Latronico}, {Lee},
  {Lemoine-Goumard}, {Lionetto}, {Llena Garde}, {Longo}, {Loparco}, {Lott},
  {Lovellette}, {Lubrano}, {Madejski}, {Makeev}, {Marangelli}, {Marelli},
  {Massaro}, {Mazziotta}, {McConville}, {McEnery}, {Michelson}, {Minuti},
  {Mitthumsiri}, {Mizuno}, {Moiseev}, {Mongelli}, {Monte}, {Monzani},
  {Moretti}, {Morselli}, {Moskalenko}, {Murgia}, {Nakajima}, {Nakamori},
  {Naumann-Godo}, {Nolan}, {Norris}, {Nuss}, {Ohno}, {Ohsugi}, {Omodei},
  {Orlando}, {Ormes}, {Ozaki}, {Paccagnella}, {Paneque}, {Panetta}, {Parent},
  {Pelassa}, {Pepe}, {Pesce-Rollins}, {Pinchera}, {Piron}, {Porter}, {Poupard},
  {Rain{\`o}}, {Rando}, {Ray}, {Razzano}, {Razzaque}, {Rea}, {Reimer},
  {Reimer}, {Reposeur}, {Ripken}, {Ritz}, {Rochester}, {Rodriguez}, {Romani},
  {Roth}, {Sadrozinski}, {Salvetti}, {Sanchez}, {Sander}, {Saz Parkinson},
  {Scargle}, {Schalk}, {Scolieri}, {Sgr{\`o}}, {Shaw}, {Siskind}, {Smith},
  {Smith}, {Spandre}, {Spinelli}, {Starck}, {Stephens}, {Striani}, {Strickman},
  {Strong}, {Suson}, {Tajima}, {Takahashi}, {Takahashi}, {Tanaka}, {Thayer},
  {Thayer}, {Thompson}, {Tibaldo}, {Tibolla}, {Tinebra}, {Torres}, {Tosti},
  {Tramacere}, {Uchiyama}, {Usher}, {Van Etten}, {Vasileiou}, {Vilchez},
  {Vitale}, {Waite}, {Wallace}, {Wang}, {Watters}, {Winer}, {Wood}, {Yang},
  {Ylinen}, \& {Ziegler}}]{2010ApJS..188..405A}
{Abdo}, A.~A., {Ackermann}, M., {Ajello}, M., {et~al.} 2010{\natexlab{a}},
  \apjs, 188, 405

\bibitem[{{Abdo} {et~al.}(2010{\natexlab{b}}){Abdo}, {Ackermann}, {Ajello},
  {Allafort}, {Antolini}, {Atwood}, {Axelsson}, {Baldini}, {Ballet},
  {Barbiellini}, {Bastieri}, {Baughman}, {Bechtol}, {Bellazzini}, {Berenji},
  {Blandford}, {Bloom}, {Bogart}, {Bonamente}, {Borgland}, {Bouvier},
  {Bregeon}, {Brez}, {Brigida}, {Bruel}, {Buehler}, {Burnett}, {Buson},
  {Caliandro}, {Cameron}, {Cannon}, {Caraveo}, {Carrigan}, {Casandjian},
  {Cavazzuti}, {Cecchi}, {{\c C}elik}, {Celotti}, {Charles}, {Chekhtman},
  {Chen}, {Cheung}, {Chiang}, {Ciprini}, {Claus}, {Cohen-Tanugi}, {Conrad},
  {Costamante}, {Cotter}, {Cutini}, {D'Elia}, {Dermer}, {de Angelis}, {de
  Palma}, {De Rosa}, {Digel}, {Silva}, {Drell}, {Dubois}, {Dumora}, {Escande},
  {Farnier}, {Favuzzi}, {Fegan}, {Ferrara}, {Focke}, {Fortin}, {Frailis},
  {Fukazawa}, {Funk}, {Fusco}, {Gargano}, {Gasparrini}, {Gehrels}, {Germani},
  {Giebels}, {Giglietto}, {Giommi}, {Giordano}, {Giroletti}, {Glanzman},
  {Godfrey}, {Grandi}, {Grenier}, {Grondin}, {Grove}, {Guiriec}, {Hadasch},
  {Harding}, {Hayashida}, {Hays}, {Healey}, {Hill}, {Horan}, {Hughes},
  {Iafrate}, {Itoh}, {J{\'o}hannesson}, {Johnson}, {Johnson}, {Johnson},
  {Johnson}, {Kamae}, {Katagiri}, {Kataoka}, {Kawai}, {Kerr}, {Kn{\"o}dlseder},
  {Kuss}, {Lande}, {Latronico}, {Lavalley}, {Lemoine-Goumard}, {Llena Garde},
  {Longo}, {Loparco}, {Lott}, {Lovellette}, {Lubrano}, {Madejski}, {Makeev},
  {Malaguti}, {Massaro}, {Mazziotta}, {McConville}, {McEnery}, {McGlynn},
  {Michelson}, {Mitthumsiri}, {Mizuno}, {Moiseev}, {Monte}, {Monzani},
  {Morselli}, {Moskalenko}, {Murgia}, {Nolan}, {Norris}, {Nuss}, {Ohno},
  {Ohsugi}, {Omodei}, {Orlando}, {Ormes}, {Ozaki}, {Paneque}, {Panetta},
  {Parent}, {Pelassa}, {Pepe}, {Pesce-Rollins}, {Piranomonte}, {Piron},
  {Porter}, {Rain{\`o}}, {Rando}, {Razzano}, {Reimer}, {Reimer}, {Reposeur},
  {Ripken}, {Ritz}, {Rodriguez}, {Romani}, {Roth}, {Ryde}, {Sadrozinski},
  {Sanchez}, {Sander}, {Saz Parkinson}, {Scargle}, {Sgr{\`o}}, {Shaw},
  {Siskind}, {Smith}, {Spandre}, {Spinelli}, {Starck}, {Stawarz}, {Strickman},
  {Suson}, {Tajima}, {Takahashi}, {Takahashi}, {Tanaka}, {Taylor}, {Thayer},
  {Thayer}, {Thompson}, {Tibaldo}, {Torres}, {Tosti}, {Tramacere}, {Ubertini},
  {Uchiyama}, {Usher}, {Vasileiou}, {Vilchez}, {Villata}, {Vitale}, {Waite},
  {Wallace}, {Wang}, {Winer}, {Wood}, {Yang}, {Ylinen}, \&
  {Ziegler}}]{2010ApJ...715..429A}
{Abdo}, A.~A., {Ackermann}, M., {Ajello}, M., {et~al.} 2010{\natexlab{b}},
  \apj, 715, 429

\bibitem[{{Abdo} {et~al.}(2010{\natexlab{c}}){Abdo}, {Ackermann}, {Ajello},
  {Antolini}, {Baldini}, {Ballet}, {Barbiellini}, {Bastieri}, {Bechtol},
  {Bellazzini}, {Berenji}, {Blandford}, {Bloom}, {Bonamente}, {Borgland},
  {Bouvier}, {Bregeon}, {Brez}, {Brigida}, {Bruel}, {Buehler}, {Burnett},
  {Buson}, {Caliandro}, {Cameron}, {Caraveo}, {Carrigan}, {Casandjian},
  {Cavazzuti}, {Cecchi}, {{\c C}elik}, {Chekhtman}, {Cheung}, {Chiang},
  {Ciprini}, {Claus}, {Cohen-Tanugi}, {Cominsky}, {Conrad}, {Costamante},
  {Cutini}, {Dermer}, {de Angelis}, {de Palma}, {Silva}, {Drell}, {Dubois},
  {Dumora}, {Farnier}, {Favuzzi}, {Fegan}, {Focke}, {Fortin}, {Frailis},
  {Fukazawa}, {Funk}, {Fusco}, {Gargano}, {Gasparrini}, {Gehrels}, {Germani},
  {Giebels}, {Giglietto}, {Giommi}, {Giordano}, {Glanzman}, {Godfrey},
  {Grenier}, {Grondin}, {Grove}, {Guiriec}, {Hadasch}, {Hayashida}, {Hays},
  {Healey}, {Horan}, {Hughes}, {Itoh}, {J{\'o}hannesson}, {Johnson}, {Johnson},
  {Kamae}, {Katagiri}, {Kataoka}, {Kawai}, {Kn{\"o}dlseder}, {Kuss}, {Lande},
  {Larsson}, {Latronico}, {Lemoine-Goumard}, {Longo}, {Loparco}, {Lott},
  {Lovellette}, {Lubrano}, {Madejski}, {Makeev}, {Massaro}, {Mazziotta},
  {McEnery}, {Michelson}, {Mitthumsiri}, {Mizuno}, {Moiseev}, {Monte},
  {Monzani}, {Morselli}, {Moskalenko}, {Mueller}, {Murgia}, {Nolan}, {Norris},
  {Nuss}, {Ohno}, {Ohsugi}, {Omodei}, {Orlando}, {Ormes}, {Ozaki}, {Panetta},
  {Parent}, {Pelassa}, {Pepe}, {Pesce-Rollins}, {Piron}, {Porter}, {Rain{\`o}},
  {Rando}, {Razzano}, {Reimer}, {Reimer}, {Ritz}, {Rodriguez}, {Romani},
  {Roth}, {Ryde}, {Sadrozinski}, {Sander}, {Scargle}, {Sgr{\`o}}, {Shaw},
  {Smith}, {Spandre}, {Spinelli}, {Starck}, {Strickman}, {Suson}, {Takahashi},
  {Takahashi}, {Tanaka}, {Thayer}, {Thayer}, {Thompson}, {Tibaldo}, {Torres},
  {Tosti}, {Tramacere}, {Uchiyama}, {Usher}, {Vasileiou}, {Vilchez}, {Vitale},
  {Waite}, {Wallace}, {Wang}, {Winer}, {Wood}, {Yang}, {Ylinen}, \&
  {Ziegler}}]{2010ApJ...722..520A}
{Abdo}, A.~A., {Ackermann}, M., {Ajello}, M., {et~al.} 2010{\natexlab{c}},
  \apj, 722, 520

\bibitem[{{Abdo} {et~al.}(2009{\natexlab{a}}){Abdo}, {Ackermann}, {Ajello},
  {Atwood}, {Axelsson}, {Baldini}, {Ballet}, {Band}, {Barbiellini}, {Bastieri},
  {Battelino}, {Baughman}, {Bechtol}, {Bellazzini}, {Berenji}, {Bignami},
  {Blandford}, {Bloom}, {Bonamente}, {Borgland}, {Bouvier}, {Bregeon}, {Brez},
  {Brigida}, {Bruel}, {Burnett}, {Caliandro}, {Cameron}, {Caraveo},
  {Casandjian}, {Cavazzuti}, {Cecchi}, {Charles}, {Chekhtman}, {Cheung},
  {Chiang}, {Ciprini}, {Claus}, {Cohen-Tanugi}, {Cominsky}, {Conrad}, {Corbet},
  {Costamante}, {Cutini}, {Davis}, {Dermer}, {de Angelis}, {de Luca}, {de
  Palma}, {Digel}, {Dormody}, {do Couto e Silva}, {Drell}, {Dubois}, {Dumora},
  {Farnier}, {Favuzzi}, {Fegan}, {Ferrara}, {Focke}, {Frailis}, {Fukazawa},
  {Funk}, {Fusco}, {Gargano}, {Gasparrini}, {Gehrels}, {Germani}, {Giebels},
  {Giglietto}, {Giommi}, {Giordano}, {Glanzman}, {Godfrey}, {Grenier},
  {Grondin}, {Grove}, {Guillemot}, {Guiriec}, {Hanabata}, {Harding}, {Hartman},
  {Hayashida}, {Hays}, {Healey}, {Horan}, {Hughes}, {J{\'o}hannesson},
  {Johnson}, {Johnson}, {Johnson}, {Johnson}, {Kamae}, {Katagiri}, {Kataoka},
  {Kawai}, {Kerr}, {Kn{\"o}dlseder}, {Kocevski}, {Kocian}, {Komin}, {Kuehn},
  {Kuss}, {Lande}, {Latronico}, {Lee}, {Lemoine-Goumard}, {Longo}, {Loparco},
  {Lott}, {Lovellette}, {Lubrano}, {Madejski}, {Makeev}, {Marelli},
  {Mazziotta}, {McConville}, {McEnery}, {McGlynn}, {Meurer}, {Michelson},
  {Mitthumsiri}, {Mizuno}, {Moiseev}, {Monte}, {Monzani}, {Moretti},
  {Morselli}, {Moskalenko}, {Murgia}, {Nakamori}, {Nolan}, {Norris}, {Nuss},
  {Ohno}, {Ohsugi}, {Omodei}, {Orlando}, {Ormes}, {Ozaki}, {Paneque},
  {Panetta}, {Parent}, {Pelassa}, {Pepe}, {Pesce-Rollins}, {Piron}, {Porter},
  {Poupard}, {Rain{\`o}}, {Rando}, {Ray}, {Razzano}, {Rea}, {Reimer}, {Reimer},
  {Reposeur}, {Ritz}, {Rochester}, {Rodriguez}, {Romani}, {Roth}, {Ryde},
  {Sadrozinski}, {Sanchez}, {Sander}, {Saz Parkinson}, {Scargle}, {Schalk},
  {Sellerholm}, {Sgr{\`o}}, {Shaw}, {Shrader}, {Sierpowska-Bartosik},
  {Siskind}, {Smith}, {Smith}, {Spandre}, {Spinelli}, {Starck}, {Stephens},
  {Strickman}, {Strong}, {Suson}, {Tajima}, {Takahashi}, {Takahashi}, {Tanaka},
  {Thayer}, {Thayer}, {Thompson}, {Tibaldo}, {Tibolla}, {Torres}, {Tosti},
  {Tramacere}, {Uchiyama}, {Usher}, {Van Etten}, {Vilchez}, {Vitale}, {Waite},
  {Wallace}, {Wang}, {Watters}, {Winer}, {Wood}, {Ylinen}, {Ziegler}, \& {The
  Fermi/LAT Collaboration}}]{2009ApJS..183...46A}
{Abdo}, A.~A., {Ackermann}, M., {Ajello}, M., {et~al.} 2009{\natexlab{a}},
  \apjs, 183, 46

\bibitem[{{Abdo} {et~al.}(2009{\natexlab{b}}){Abdo}, {Ackermann}, {Ajello},
  {Atwood}, {Axelsson}, {Baldini}, {Ballet}, {Barbiellini}, {Bastieri},
  {Baughman}, {Bechtol}, {Bellazzini}, {Blandford}, {Bloom}, {Bonamente},
  {Borgland}, {Bouvier}, {Bregeon}, {Brez}, {Brigida}, {Bruel}, {Burnett},
  {Caliandro}, {Cameron}, {Caraveo}, {Casandjian}, {Cavazzuti}, {Cecchi},
  {Charles}, {Chekhtman}, {Chen}, {Cheung}, {Chiang}, {Ciprini}, {Claus},
  {Cohen-Tanugi}, {Colafrancesco}, {Collmar}, {Cominsky}, {Conrad},
  {Costamante}, {Cutini}, {Dermer}, {de Angelis}, {de Palma}, {Digel}, {do
  Couto e Silva}, {Drell}, {Dubois}, {Dumora}, {Farnier}, {Favuzzi}, {Fegan},
  {Ferrara}, {Finke}, {Focke}, {Foschini}, {Frailis}, {Fuhrmann}, {Fukazawa},
  {Funk}, {Fusco}, {Gargano}, {Gasparrini}, {Gehrels}, {Germani}, {Giebels},
  {Giglietto}, {Giommi}, {Giordano}, {Giroletti}, {Glanzman}, {Godfrey},
  {Grenier}, {Grondin}, {Grove}, {Guillemot}, {Guiriec}, {Hanabata}, {Harding},
  {Hartman}, {Hayashida}, {Hays}, {Healey}, {Horan}, {Hughes},
  {J{\'o}hannesson}, {Johnson}, {Johnson}, {Johnson}, {Johnson}, {Kadler},
  {Kamae}, {Katagiri}, {Kataoka}, {Kerr}, {Kn{\"o}dlseder}, {Kocian}, {Kuehn},
  {Kuss}, {Lande}, {Latronico}, {Lemoine-Goumard}, {Longo}, {Loparco}, {Lott},
  {Lovellette}, {Lubrano}, {Madejski}, {Makeev}, {Massaro}, {Mazziotta},
  {McConville}, {McEnery}, {McGlynn}, {Meurer}, {Michelson}, {Mitthumsiri},
  {Mizuno}, {Moiseev}, {Monte}, {Monzani}, {Moretti}, {Morselli}, {Moskalenko},
  {Murgia}, {Nolan}, {Norris}, {Nuss}, {Ohsugi}, {Omodei}, {Orlando}, {Ormes},
  {Ozaki}, {Paneque}, {Panetta}, {Parent}, {Pelassa}, {Pepe}, {Pesce-Rollins},
  {Piron}, {Porter}, {Rain{\`o}}, {Rando}, {Razzano}, {Razzaque}, {Reimer},
  {Reimer}, {Reposeur}, {Reyes}, {Ritz}, {Rochester}, {Rodriguez}, {Romani},
  {Ryde}, {Sadrozinski}, {Sanchez}, {Sander}, {Saz Parkinson}, {Scargle},
  {Schalk}, {Sellerholm}, {Sgr{\`o}}, {Shaw}, {Smith}, {Smith}, {Spandre},
  {Spinelli}, {Starck}, {Strickman}, {Suson}, {Tajima}, {Takahashi},
  {Takahashi}, {Tanaka}, {Taylor}, {Thayer}, {Thayer}, {Thompson}, {Tibaldo},
  {Torres}, {Tosti}, {Tramacere}, {Uchiyama}, {Usher}, {Vilchez}, {Villata},
  {Vitale}, {Waite}, {Winer}, {Wood}, {Ylinen}, \&
  {Ziegler}}]{2009ApJ...700..597A}
{Abdo}, A.~A., {Ackermann}, M., {Ajello}, M., {et~al.} 2009{\natexlab{b}},
  \apj, 700, 597

\bibitem[{{Abdo} {et~al.}(2009{\natexlab{c}}){Abdo}, {Ackermann}, {Ajello},
  {Atwood}, {Baldini}, {Ballet}, {Barbiellini}, {Bastieri}, {Baughman},
  {Bechtol}, {Bellazzini}, {Berenji}, {Bloom}, {Bonamente}, {Borgland},
  {Bouvier}, {Bregeon}, {Brez}, {Brigida}, {Bruel}, {Buehler}, {Burnett},
  {Buson}, {Caliandro}, {Cameron}, {Caraveo}, \&
  {Casandjian}}]{PhysRevD.80.122004}
{Abdo}, A.~A., {Ackermann}, M., {Ajello}, M., {et~al.} 2009{\natexlab{c}},
  Phys. Rev. D, 80, 122004

\bibitem[{{Ackermann} {et~al.}(2010){Ackermann}, {Ajello}, {Baldini}, {Ballet},
  {Barbiellini}, {Bastieri}, {Bechtol}, {Bellazzini}, {Berenji}, {Blandford},
  {Bonamente}, {Borgland}, {Bregeon}, {Brigida}, {Bruel}, {Buehler}, {Burnett},
  {Buson}, {Caliandro}, {Cameron}, {Caraveo}, {Carrigan}, {Casandjian},
  {Cavazzuti}, {Cecchi}, {{\c C}elik}, {Chekhtman}, {Cheung}, {Chiang},
  {Ciprini}, {Claus}, {Cohen-Tanugi}, {Corbel}, {Cutini}, {D'Ammando},
  {Dermer}, {de Angelis}, {de Palma}, {Digel}, {Silva}, {Drell}, {Dubois},
  {Dumora}, {Escande}, {Favuzzi}, {Fegan}, {Ferrara}, {Fuhrmann}, {Fukazawa},
  {Fusco}, {Gargano}, {Gasparrini}, {Gehrels}, {Germani}, {Giebels},
  {Giglietto}, {Giommi}, {Giordano}, {Giroletti}, {Glanzman}, {Godfrey},
  {Grenier}, {Grove}, {Guiriec}, {Hadasch}, {Hayashida}, {Hays},
  {J{\'o}hannesson}, {Johnson}, {Johnson}, {Kamae}, {Katagiri}, {Kataoka},
  {Kn{\"o}dlseder}, {Kuss}, {Lande}, {Larsson}, {Latronico}, {Lee}, {Llena
  Garde}, {Longo}, {Loparco}, {Lott}, {Lubrano}, {Madejski}, {Makeev},
  {Marchili}, {Mazziotta}, {McEnery}, {Mehault}, {Michelson}, {Mizuno},
  {Monte}, {Monzani}, {Morselli}, {Moskalenko}, {Murgia}, {Nakamori},
  {Nalewajko}, {Naumann-Godo}, {Nolan}, {Norris}, {Nuss}, {Ohsugi}, {Okumura},
  {Omodei}, {Orlando}, {Ormes}, {Pelassa}, {Pepe}, {Pesce-Rollins}, {Piron},
  {Porter}, {Rain{\`o}}, {Rando}, {Razzano}, {Reimer}, {Reimer}, {Reyes},
  {Ripken}, {Ritz}, {Roth}, {Sadrozinski}, {Sanchez}, {Sander}, {Scargle},
  {Sgr{\`o}}, {Sikora}, {Siskind}, {Spandre}, {Spinelli}, {Strickman}, {Suson},
  {Takahashi}, {Takahashi}, {Tanaka}, {Tanaka}, {Thayer}, {Thayer}, {Thompson},
  {Tibaldo}, {Torres}, {Tosti}, {Tramacere}, {Usher}, {Vandenbroucke},
  {Vilchez}, {Vitale}, {Waite}, {Wang}, {Wehrle}, {Winer}, {Yang}, {Ylinen}, \&
  {Ziegler}}]{2010ApJ...721.1383A}
{Ackermann}, M., {Ajello}, M., {Baldini}, L., {et~al.} 2010, \apj, 721, 1383

\bibitem[{{Agudo} {et~al.}(2011{\natexlab{a}}){Agudo}, {Jorstad}, {Marscher},
  {Larionov}, {G{\'o}mez}, {L{\"a}hteenm{\"a}ki}, {Gurwell}, {Smith},
  {Wiesemeyer}, {Thum}, {Heidt}, {Blinov}, {D'Arcangelo}, {Hagen-Thorn},
  {Morozova}, {Nieppola}, {Roca-Sogorb}, {Schmidt}, {Taylor}, {Tornikoski}, \&
  {Troitsky}}]{2011ApJ...726L..13A}
{Agudo}, I., {Jorstad}, S.~G., {Marscher}, A.~P., {et~al.} 2011{\natexlab{a}},
  \apjl, 726, L13

\bibitem[{{Agudo} {et~al.}(2011{\natexlab{b}}){Agudo}, {Marscher}, {Jorstad},
  {Larionov}, {Gomez}, {Lahteenmaki}, {Smith}, {Nilsson}, {Readhead}, {Aller},
  {Heidt}, {Gurwell}, {Thum}, {Wehrle}, {Nikolashvili}, {Aller}, {Benitez},
  {Blinov}, {Hagen-Thorn}, {Hiriart}, {Jannuzi}, {Joshi}, {Kimeridze},
  {Kurtanidze}, {Kurtanidze}, {Lindfors}, {Molina}, {Morozova}, {Nieppola},
  {Olmstead}, {Reinthal}, {Roca-Sogorb}, {Schmidt}, {Sigua}, {Sillanpaa},
  {Takalo}, {Taylor}, {Tornikoski}, {Troitsky}, {Zook}, \&
  {Wiesemeyer}}]{2011arXiv1105.0549A}
{Agudo}, I., {Marscher}, A.~P., {Jorstad}, S.~G., {et~al.} 2011{\natexlab{b}},
  \apjl, 735, L10

\bibitem[{{Atwood} {et~al.}(2009){Atwood}, {Abdo}, {Ackermann}, {Althouse},
  {Anderson}, {Axelsson}, {Baldini}, {Ballet}, {Band}, {Barbiellini},
  {Bartelt}, {Bastieri}, {Baughman}, {Bechtol}, {B{\'e}d{\'e}r{\`e}de},
  {Bellardi}, {Bellazzini}, {Berenji}, {Bignami}, {Bisello}, {Bissaldi},
  {Blandford}, {Bloom}, {Bogart}, {Bonamente}, {Bonnell}, {Borgland},
  {Bouvier}, {Bregeon}, {Brez}, {Brigida}, {Bruel}, {Burnett}, {Busetto},
  {Caliandro}, {Cameron}, {Caraveo}, {Carius}, {Carlson}, {Casandjian},
  {Cavazzuti}, {Ceccanti}, {Cecchi}, {Charles}, {Chekhtman}, {Cheung},
  {Chiang}, {Chipaux}, {Cillis}, {Ciprini}, {Claus}, {Cohen-Tanugi},
  {Condamoor}, {Conrad}, {Corbet}, {Corucci}, {Costamante}, {Cutini}, {Davis},
  {Decotigny}, {DeKlotz}, {Dermer}, {de Angelis}, {Digel}, {do Couto e Silva},
  {Drell}, {Dubois}, {Dumora}, {Edmonds}, {Fabiani}, {Farnier}, {Favuzzi},
  {Flath}, {Fleury}, {Focke}, {Funk}, {Fusco}, {Gargano}, {Gasparrini},
  {Gehrels}, {Gentit}, {Germani}, {Giebels}, {Giglietto}, {Giommi}, {Giordano},
  {Glanzman}, {Godfrey}, {Grenier}, {Grondin}, {Grove}, {Guillemot}, {Guiriec},
  {Haller}, {Harding}, {Hart}, {Hays}, {Healey}, {Hirayama}, {Hjalmarsdotter},
  {Horn}, {Hughes}, {J{\'o}hannesson}, {Johansson}, {Johnson}, {Johnson},
  {Johnson}, {Johnson}, {Kamae}, {Katagiri}, {Kataoka}, {Kavelaars}, {Kawai},
  {Kelly}, {Kerr}, {Klamra}, {Kn{\"o}dlseder}, {Kocian}, {Komin}, {Kuehn},
  {Kuss}, {Landriu}, {Latronico}, {Lee}, {Lee}, {Lemoine-Goumard}, {Lionetto},
  {Longo}, {Loparco}, {Lott}, {Lovellette}, {Lubrano}, {Madejski}, {Makeev},
  {Marangelli}, {Massai}, {Mazziotta}, {McEnery}, {Menon}, {Meurer},
  {Michelson}, {Minuti}, {Mirizzi}, {Mitthumsiri}, {Mizuno}, {Moiseev},
  {Monte}, {Monzani}, {Moretti}, {Morselli}, {Moskalenko}, {Murgia},
  {Nakamori}, {Nishino}, {Nolan}, {Norris}, {Nuss}, {Ohno}, {Ohsugi}, {Omodei},
  {Orlando}, {Ormes}, {Paccagnella}, {Paneque}, {Panetta}, {Parent}, {Pearce},
  {Pepe}, {Perazzo}, {Pesce-Rollins}, {Picozza}, {Pieri}, {Pinchera}, {Piron},
  {Porter}, {Poupard}, {Rain{\`o}}, {Rando}, {Rapposelli}, {Razzano}, {Reimer},
  {Reimer}, {Reposeur}, {Reyes}, {Ritz}, {Rochester}, {Rodriguez}, {Romani},
  {Roth}, {Russell}, {Ryde}, {Sabatini}, {Sadrozinski}, {Sanchez}, {Sander},
  {Sapozhnikov}, {Parkinson}, {Scargle}, {Schalk}, {Scolieri}, {Sgr{\`o}},
  {Share}, {Shaw}, {Shimokawabe}, {Shrader}, {Sierpowska-Bartosik}, {Siskind},
  {Smith}, {Smith}, {Spandre}, {Spinelli}, {Starck}, {Stephens}, {Strickman},
  {Strong}, {Suson}, {Tajima}, {Takahashi}, {Takahashi}, {Tanaka}, {Tenze},
  {Tether}, {Thayer}, {Thayer}, {Thompson}, {Tibaldo}, {Tibolla}, {Torres},
  {Tosti}, {Tramacere}, {Turri}, {Usher}, {Vilchez}, {Vitale}, {Wang},
  {Watters}, {Winer}, {Wood}, {Ylinen}, \& {Ziegler}}]{2009ApJ...697.1071A}
{Atwood}, W.~B., {Abdo}, A.~A., {Ackermann}, M., {et~al.} 2009, \apj, 697, 1071

\bibitem[{{Bessell} {et~al.}(1998){Bessell}, {Castelli}, \&
  {Plez}}]{1998A&A...333..231B}
{Bessell}, M.~S., {Castelli}, F., \& {Plez}, B. 1998, \aap, 333, 231

\bibitem[{{Brinkmann} {et~al.}(1995){Brinkmann}, {Siebert}, {Reich}, {Fuerst},
  {Reich}, {Voges}, {Truemper}, \& {Wielebinski}}]{1995A&AS..109..147B}
{Brinkmann}, W., {Siebert}, J., {Reich}, W., {et~al.} 1995, \aaps, 109, 147

\bibitem[{{Burrows} {et~al.}(2005){Burrows}, {Hill}, {Nousek}, {Kennea},
  {Wells}, {Osborne}, {Abbey}, {Beardmore}, {Mukerjee}, {Short}, {Chincarini},
  {Campana}, {Citterio}, {Moretti}, {Pagani}, {Tagliaferri}, {Giommi},
  {Capalbi}, {Tamburelli}, {Angelini}, {Cusumano}, {Br{\"a}uninger}, {Burkert},
  \& {Hartner}}]{2005SSRv..120..165B}
{Burrows}, D.~N., {Hill}, J.~E., {Nousek}, J.~A., {et~al.} 2005, Space Science
  Reviews, 120, 165

\bibitem[{{Bychkova} {et~al.}(2004){Bychkova}, {Kardashev}, {Vlasyuk}, \&
  {Spiridonova}}]{2004ARep...48..840B}
{Bychkova}, V.~S., {Kardashev}, N.~S., {Vlasyuk}, V.~V., \& {Spiridonova},
  O.~I. 2004, Astronomy Reports, 48, 840

\bibitem[{{Cardelli} {et~al.}(1989){Cardelli}, {Clayton}, \&
  {Mathis}}]{1989ApJ...345..245C}
{Cardelli}, J.~A., {Clayton}, G.~C., \& {Mathis}, J.~S. 1989, \apj, 345, 245

\bibitem[{{Casandjian} \& {Grenier}(2008)}]{2008A&A...489..849C}
{Casandjian}, J.-M. \& {Grenier}, I.~A. 2008, \aap, 489, 849

\bibitem[{{Cash}(1979)}]{1979ApJ...228..939C}
{Cash}, W. 1979, \apj, 228, 939

\bibitem[{{Chatterjee} {et~al.}(2011){Chatterjee}, {Bailyn}, {Bonning},
  {Buxton}, {Coppi}, {Isler}, \& {Urry}}]{2011arXiv1101.3815C}
{Chatterjee}, R., {Bailyn}, C., {Bonning}, E.~W., {et~al.} 2011, ArXiv
  e-prints, arXiv:1101.3815

\bibitem[{{Chatterjee} {et~al.}(2008){Chatterjee}, {Jorstad}, {Marscher}, {Oh},
  {McHardy}, {Aller}, {Aller}, {Balonek}, {Miller}, {Ryle}, {Tosti},
  {Kurtanidze}, {Nikolashvili}, {Larionov}, \&
  {Hagen-Thorn}}]{2008ApJ...689...79C}
{Chatterjee}, R., {Jorstad}, S.~G., {Marscher}, A.~P., {et~al.} 2008, \apj,
  689, 79

\bibitem[{{Edelson} \& {Krolik}(1988)}]{1988ApJ...333..646E}
{Edelson}, R.~A. \& {Krolik}, J.~H. 1988, \apj, 333, 646

\bibitem[{{Fey} {et~al.}(2004){Fey}, {Ma}, {Arias}, {Charlot},
  {Feissel-Vernier}, {Gontier}, {Jacobs}, {Li}, \&
  {MacMillan}}]{2004AJ....127.3587F}
{Fey}, A.~L., {Ma}, C., {Arias}, E.~F., {et~al.} 2004, \aj, 127, 3587

\bibitem[{{Fichtel} {et~al.}(1994){Fichtel}, {Bertsch}, {Chiang}, {Dingus},
  {Esposito}, {Fierro}, {Hartman}, {Hunter}, {Kanbach}, {Kniffen}, {Kwok},
  {Lin}, {Mattox}, {Mayer-Hasselwander}, {McDonald}, {Michelson}, {von
  Montigny}, {Nolan}, {Pinkau}, {Radecke}, {Rothermel}, {Sreekumar}, {Sommer},
  {Schneid}, {Thompson}, \& {Willis}}]{1994ApJS...94..551F}
{Fichtel}, C.~E., {Bertsch}, D.~L., {Chiang}, J., {et~al.} 1994, \apjs, 94, 551

\bibitem[{{Gehrels} {et~al.}(2004){Gehrels}, {Chincarini}, {Giommi}, {Mason},
  {Nousek}, {Wells}, {White}, {Barthelmy}, {Burrows}, {Cominsky}, {Hurley},
  {Marshall}, {M{\'e}sz{\'a}ros}, {Roming}, {Angelini}, {Barbier}, {Belloni},
  {Campana}, {Caraveo}, {Chester}, {Citterio}, {Cline}, {Cropper}, {Cummings},
  {Dean}, {Feigelson}, {Fenimore}, {Frail}, {Fruchter}, {Garmire}, {Gendreau},
  {Ghisellini}, {Greiner}, {Hill}, {Hunsberger}, {Krimm}, {Kulkarni}, {Kumar},
  {Lebrun}, {Lloyd-Ronning}, {Markwardt}, {Mattson}, {Mushotzky}, {Norris},
  {Osborne}, {Paczynski}, {Palmer}, {Park}, {Parsons}, {Paul}, {Rees},
  {Reynolds}, {Rhoads}, {Sasseen}, {Schaefer}, {Short}, {Smale}, {Smith},
  {Stella}, {Tagliaferri}, {Takahashi}, {Tashiro}, {Townsley}, {Tueller},
  {Turner}, {Vietri}, {Voges}, {Ward}, {Willingale}, {Zerbi}, \&
  {Zhang}}]{2004ApJ...611.1005G}
{Gehrels}, N., {Chincarini}, G., {Giommi}, P., {et~al.} 2004, \apj, 611, 1005

\bibitem[{{Gonz{\'a}lez-P{\'e}rez} {et~al.}(2001){Gonz{\'a}lez-P{\'e}rez},
  {Kidger}, \& {Mart{\'{\i}}n-Luis}}]{2001AJ....122.2055G}
{Gonz{\'a}lez-P{\'e}rez}, J.~N., {Kidger}, M.~R., \& {Mart{\'{\i}}n-Luis}, F.
  2001, \aj, 122, 2055

\bibitem[{{Hartman} {et~al.}(1999){Hartman}, {Bertsch}, {Bloom}, {Chen},
  {Deines-Jones}, {Esposito}, {Fichtel}, {Friedlander}, {Hunter}, {McDonald},
  {Sreekumar}, {Thompson}, {Jones}, {Lin}, {Michelson}, {Nolan}, {Tompkins},
  {Kanbach}, {Mayer-Hasselwander}, {M{\"u}cke}, {Pohl}, {Reimer}, {Kniffen},
  {Schneid}, {von Montigny}, {Mukherjee}, \& {Dingus}}]{1999ApJS..123...79H}
{Hartman}, R.~C., {Bertsch}, D.~L., {Bloom}, S.~D., {et~al.} 1999, \apjs, 123,
  79

\bibitem[{{Kalberla} {et~al.}(2005){Kalberla}, {Burton}, {Hartmann}, {Arnal},
  {Bajaja}, {Morras}, \& {P{\"o}ppel}}]{2005A&A...440..775K}
{Kalberla}, P.~M.~W., {Burton}, W.~B., {Hartmann}, D., {et~al.} 2005, \aap,
  440, 775

\bibitem[{{Kataoka} {et~al.}(1999){Kataoka}, {Mattox}, {Quinn}, {Kubo},
  {Makino}, {Takahashi}, {Inoue}, {Hartman}, {Madejski}, {Sreekumar}, \&
  {Wagner}}]{1999ApJ...514..138K}
{Kataoka}, J., {Mattox}, J.~R., {Quinn}, J., {et~al.} 1999, \apj, 514, 138

\bibitem[{{Kollgaard} {et~al.}(1989){Kollgaard}, {Wardle}, \&
  {Roberts}}]{1989AJ.....97.1550K}
{Kollgaard}, R.~I., {Wardle}, J.~F.~C., \& {Roberts}, D.~H. 1989, \aj, 97, 1550

\bibitem[{{Kovalev} {et~al.}(1999){Kovalev}, {Nizhelsky}, {Kovalev}, {Berlin},
  {Zhekanis}, {Mingaliev}, \& {Bogdantsov}}]{1999yCat..41390545K}
{Kovalev}, Y.~Y., {Nizhelsky}, N.~A., {Kovalev}, Y.~A., {et~al.} 1999, VizieR
  Online Data Catalog, 413, 90545

\bibitem[{{Larionov} {et~al.}(2009){Larionov}, {Villata}, {Raiteri},
  {Carosati}, {Ros}, {Casas}, {Bravo}, {Melnichuk}, \&
  {Gurwell}}]{2009ATel.2222....1L}
{Larionov}, V.~M., {Villata}, M., {Raiteri}, C.~M., {et~al.} 2009, The
  Astronomer's Telegram, 2222, 1

\bibitem[{{Lister} {et~al.}(2009){Lister}, {Aller}, {Aller}, {Cohen}, {Homan},
  {Kadler}, {Kellermann}, {Kovalev}, {Ros}, {Savolainen}, {Zensus}, \&
  {Vermeulen}}]{2009AJ....137.3718L}
{Lister}, M.~L., {Aller}, H.~D., {Aller}, M.~F., {et~al.} 2009, \aj, 137, 3718

\bibitem[{{Lobanov} \& {Zensus}(1999)}]{1999ApJ...521..509L}
{Lobanov}, A.~P. \& {Zensus}, J.~A. 1999, \apj, 521, 509

\bibitem[{{Marziani} {et~al.}(1996){Marziani}, {Sulentic}, {Dultzin-Hacyan},
  {Calvani}, \& {Moles}}]{1996ApJS..104...37M}
{Marziani}, P., {Sulentic}, J.~W., {Dultzin-Hacyan}, D., {Calvani}, M., \&
  {Moles}, M. 1996, \apjs, 104, 37

\bibitem[{{Mattox} {et~al.}(1996){Mattox}, {Bertsch}, {Chiang}, {Dingus},
  {Digel}, {Esposito}, {Fierro}, {Hartman}, {Hunter}, {Kanbach}, {Kniffen},
  {Lin}, {Macomb}, {Mayer-Hasselwander}, {Michelson}, {von Montigny},
  {Mukherjee}, {Nolan}, {Ramanamurthy}, {Schneid}, {Sreekumar}, {Thompson}, \&
  {Willis}}]{1996ApJ...461..396M}
{Mattox}, J.~R., {Bertsch}, D.~L., {Chiang}, J., {et~al.} 1996, \apj, 461, 396

\bibitem[{{Max-Moerbeck} {et~al.}(2010){Max-Moerbeck}, {Richards}, {Pavlidou},
  {Pearson}, {Readhead}, {Stevenson}, {King}, {Reeves}, {Karkare}, {Angelakis},
  {Fuhrmann}, {Zensus}, {Healey}, {Romani}, \& {Shaw}}]{FmJ2010}
{Max-Moerbeck}, W., {Richards}, J.~L., {Pavlidou}, V., {et~al.} 2010, in Fermi
  meets Jansky - AGN in Radio and Gamma-Rays, ed. T.~{Savolainen}, E.~{Ros},
  W.~P. {Porcas}, \& J.~A. {Zensus}, 77--80

\bibitem[{{Myers} {et~al.}(2003){Myers}, {Jackson}, {Browne}, {de Bruyn},
  {Pearson}, {Readhead}, {Wilkinson}, {Biggs}, {Blandford}, {Fassnacht},
  {Koopmans}, {Marlow}, {McKean}, {Norbury}, {Phillips}, {Rusin}, {Shepherd},
  \& {Sykes}}]{2003MNRAS.341....1M}
{Myers}, S.~T., {Jackson}, N.~J., {Browne}, I.~W.~A., {et~al.} 2003, \mnras,
  341, 1

\bibitem[{{Poole} {et~al.}(2008){Poole}, {Breeveld}, {Page}, {Landsman},
  {Holland}, {Roming}, {Kuin}, {Brown}, {Gronwall}, {Hunsberger}, {Koch},
  {Mason}, {Schady}, {vanden Berk}, {Blustin}, {Boyd}, {Broos}, {Carter},
  {Chester}, {Cucchiara}, {Hancock}, {Huckle}, {Immler}, {Ivanushkina},
  {Kennedy}, {Marshall}, {Morgan}, {Pandey}, {de Pasquale}, {Smith}, \&
  {Still}}]{2008MNRAS.383..627P}
{Poole}, T.~S., {Breeveld}, A.~A., {Page}, M.~J., {et~al.} 2008, \mnras, 383,
  627

\bibitem[{{Pushkarev} {et~al.}(2010){Pushkarev}, {Kovalev}, \&
  {Lister}}]{2010ApJ...722L...7P}
{Pushkarev}, A.~B., {Kovalev}, Y.~Y., \& {Lister}, M.~L. 2010, \apjl, 722, L7

\bibitem[{{Pushkarev} {et~al.}(2009){Pushkarev}, {Kovalev}, {Lister}, \&
  {Savolainen}}]{2009A&A...507L..33P}
{Pushkarev}, A.~B., {Kovalev}, Y.~Y., {Lister}, M.~L., \& {Savolainen}, T.
  2009, \aap, 507, L33

\bibitem[{{Rando et al.}(2009)}]{2009arXiv0907.0626R}
{Rando et al.} 2009, ArXiv e-prints, arXiv:0907.0626

\bibitem[{{Reyes} \& {Cheung}(2009)}]{2009ATel.2226....1R}
{Reyes}, L.~C. \& {Cheung}, C.~C. 2009, The Astronomer's Telegram, 2226, 1

\bibitem[{{Rolke} {et~al.}(2005){Rolke}, {L\'opez}, \&
  {Conrad}}]{2005NIMA.551.493...R}
{Rolke}, W.~A., {L\'opez}, A.~M., \& {Conrad}, J. 2005, NIM A, 551, 493

\bibitem[{{Roming} {et~al.}(2005){Roming}, {Kennedy}, {Mason}, {Nousek}, {Ahr},
  {Bingham}, {Broos}, {Carter}, {Hancock}, {Huckle}, {Hunsberger}, {Kawakami},
  {Killough}, {Koch}, {McLelland}, {Smith}, {Smith}, {Soto}, {Boyd},
  {Breeveld}, {Holland}, {Ivanushkina}, {Pryzby}, {Still}, \&
  {Stock}}]{2005SSRv..120...95R}
{Roming}, P.~W.~A., {Kennedy}, T.~E., {Mason}, K.~O., {et~al.} 2005, Space
  Science Reviews, 120, 95

\bibitem[{{Roming} {et~al.}(2009){Roming}, {Koch}, {Oates}, {Porterfield},
  {Vanden Berk}, {Boyd}, {Holland}, {Hoversten}, {Immler}, {Marshall}, {Page},
  {Racusin}, {Schneider}, {Breeveld}, {Brown}, {Chester}, {Cucchiara},
  {DePasquale}, {Gronwall}, {Hunsberger}, {Kuin}, {Landsman}, {Schady}, \&
  {Still}}]{2009ApJ...690..163R}
{Roming}, P.~W.~A., {Koch}, T.~S., {Oates}, S.~R., {et~al.} 2009, \apj, 690,
  163

\bibitem[{{Schinzel} {et~al.}(2010){Schinzel}, {Lobanov}, {Jorstad},
  {Marscher}, {Taylor}, \& {Zensus}}]{SchinzelFmJ2010}
{Schinzel}, F.~K., {Lobanov}, A.~P., {Jorstad}, S.~G., {et~al.} 2010, in Fermi
  meets Jansky - AGN in Radio and Gamma-Rays, ed. T.~{Savolainen}, E.~{Ros},
  W.~P. {Porcas}, \& J.~A. {Zensus}, 175--178, arXiv:1012.2820

\bibitem[{{Schlegel} {et~al.}(1998){Schlegel}, {Finkbeiner}, \&
  {Davis}}]{1998ApJ...500..525S}
{Schlegel}, D.~J., {Finkbeiner}, D.~P., \& {Davis}, M. 1998, \apj, 500, 525

\bibitem[{{Smith} {et~al.}(2002){Smith}, {Tucker}, {Kent}, {Richmond},
  {Fukugita}, {Ichikawa}, {Ichikawa}, {Jorgensen}, {Uomoto}, {Gunn}, {Hamabe},
  {Watanabe}, {Tolea}, {Henden}, {Annis}, {Pier}, {McKay}, {Brinkmann}, {Chen},
  {Holtzman}, {Shimasaku}, \& {York}}]{2002AJ....123.2121S}
{Smith}, J.~A., {Tucker}, D.~L., {Kent}, S., {et~al.} 2002, \aj, 123, 2121

\bibitem[{{Smith} {et~al.}(1985){Smith}, {Balonek}, {Heckert}, {Elston}, \&
  {Schmidt}}]{1985AJ.....90.1184S}
{Smith}, P.~S., {Balonek}, T.~J., {Heckert}, P.~A., {Elston}, R., \& {Schmidt},
  G.~D. 1985, \aj, 90, 1184

\bibitem[{{Stickel} {et~al.}(1989){Stickel}, {Fried}, \&
  {Kuehr}}]{1989A&AS...80..103S}
{Stickel}, M., {Fried}, J.~W., \& {Kuehr}, H. 1989, \aaps, 80, 103

\bibitem[{{Tavecchio} {et~al.}(2010){Tavecchio}, {Ghisellini}, {Bonnoli}, \&
  {Ghirlanda}}]{2010MNRAS.405L..94T}
{Tavecchio}, F., {Ghisellini}, G., {Bonnoli}, G., \& {Ghirlanda}, G. 2010,
  \mnras, 405, L94

\bibitem[{{Thompson} {et~al.}(1993){Thompson}, {Bertsch}, {Fichtel}, {Hartman},
  {Hofstadter}, {Hughes}, {Hunter}, {Hughlock}, {Kanbach}, {Kniffen}, {Lin},
  {Mattox}, {Mayer-Hasselwander}, {von Montigny}, {Nolan}, {Nel}, {Pinkau},
  {Rothermel}, {Schneid}, {Sommer}, {Sreekumar}, {Tieger}, \&
  {Walker}}]{1993ApJS...86..629T}
{Thompson}, D.~J., {Bertsch}, D.~L., {Fichtel}, C.~E., {et~al.} 1993, \apjs,
  86, 629

\bibitem[{{Timmer} \& {Koenig}(1995)}]{1995A&A...300..707T}
{Timmer}, J. \& {Koenig}, M. 1995, \aap, 300, 707

\bibitem[{{Unwin} {et~al.}(1997){Unwin}, {Wehrle}, {Lobanov}, {Zensus},
  {Madejski}, {Aller}, \& {Aller}}]{1997ApJ...480..596U}
{Unwin}, S.~C., {Wehrle}, A.~E., {Lobanov}, A.~P., {et~al.} 1997, \apj, 480,
  596

\bibitem[{{Unwin} {et~al.}(1994){Unwin}, {Wehrle}, {Urry}, {Gilmore}, {Barton},
  {Kjerulf}, {Zensus}, \& {Rabaca}}]{1994ApJ...432..103U}
{Unwin}, S.~C., {Wehrle}, A.~E., {Urry}, C.~M., {et~al.} 1994, \apj, 432, 103

\bibitem[{{Uttley} {et~al.}(2002){Uttley}, {McHardy}, \&
  {Papadakis}}]{2002MNRAS.332..231U}
{Uttley}, P., {McHardy}, I.~M., \& {Papadakis}, I.~E. 2002, \mnras, 332, 231

\bibitem[{{Villata} {et~al.}(2009){Villata}, {Raiteri}, {Gurwell}, {Larionov},
  {Kurtanidze}, {Aller}, {L{\"a}hteenm{\"a}ki}, {Chen}, {Nilsson}, {Agudo},
  {Aller}, {Arkharov}, {Bach}, {Bachev}, {Beltrame}, {Ben{\'{\i}}tez}, {Buemi},
  {B{\"o}ttcher}, {Calcidese}, {Capezzali}, {Carosati}, {da Rio}, {di Paola},
  {Dolci}, {Dultzin}, {Forn{\'e}}, {G{\'o}mez}, {Hagen-Thorn}, {Halkola},
  {Heidt}, {Hiriart}, {Hovatta}, {Hsiao}, {Jorstad}, {Kimeridze},
  {Konstantinova}, {Kopatskaya}, {Koptelova}, {Leto}, {Ligustri}, {Lindfors},
  {Lopez}, {Marscher}, {Mommert}, {Mujica}, {Nikolashvili}, {Palma}, {Pasanen},
  {Roca-Sogorb}, {Ros}, {Roustazadeh}, {Sadun}, {Saino}, {Sigua}, {Sorcia},
  {Takalo}, {Tornikoski}, {Trigilio}, {Turchetti}, \&
  {Umana}}]{2009A&A...504L...9V}
{Villata}, M., {Raiteri}, C.~M., {Gurwell}, M.~A., {et~al.} 2009, \aap, 504, L9

\bibitem[{{Villata} {et~al.}(2008){Villata}, {Raiteri}, {Larionov},
  {Kurtanidze}, {Nilsson}, {Aller}, {Tornikoski}, {Volvach}, {Aller},
  {Arkharov}, {Bach}, {Beltrame}, {Bhatta}, {Buemi}, {B{\"o}ttcher},
  {Calcidese}, {Carosati}, {Castro-Tirado}, {da Rio}, {di Paola}, {Dolci},
  {Forn{\'e}}, {Frasca}, {Hagen-Thorn}, {Heidt}, {Hiriart}, {Jel{\'{\i}}nek},
  {Kimeridze}, {Konstantinova}, {Kopatskaya}, {Lanteri}, {Leto}, {Ligustri},
  {Lindfors}, {L{\"a}hteenm{\"a}ki}, {Marilli}, {Nieppola}, {Nikolashvili},
  {Pasanen}, {Ragozzine}, {Ros}, {Sigua}, {Smart}, {Sorcia}, {Takalo},
  {Tavani}, {Trigilio}, {Turchetti}, {Uckert}, {Umana}, {Vercellone}, \&
  {Webb}}]{2008A&A...481L..79V}
{Villata}, M., {Raiteri}, C.~M., {Larionov}, V.~M., {et~al.} 2008, \aap, 481,
  L79

\bibitem[{{Wilks}(1938)}]{1938AMS...9...60W}
{Wilks}, S.~S. 1938, Ann. Math. Stat., 9, 60

\end{thebibliography}

\end{document}